# Design Principles for Tailoring Heat Transport via Iris-Gated Core–Double–Shell Nanoparticles in the Context of Photothermal Therapies

*Javier González-Colsa[a,1,]*, Fernando Bresme[b], Pablo Albella[a,]**

E-mail: javier.gonzalez.colsa@uvigo.gal

E-mail: pablo.albella@unican.es

[a]Group of Optics, Department of Applied Physics, University of Cantabria, 39005 Santander, Spain.

[b]Department of Chemistry, Molecular Sciences Research Hub, Imperial College London, London W12 0BZ, UK.

Present address: [1]CINBIO, University of Vigo, Campus Universitario de Vigo, Lagoas Marcosede, Vigo 36310, Spain.

**Abstract**

The rational design of Janus nanostructures that combine efficient optical absorption with controlled thermal transport is essential for advancing plasmonic photothermal therapies and related applications. Here, we introduce a theoretical and computational framework to investigate polymer functionalized core–double–shell nanoparticles and their asymmetric version, the iris-gated core–double–shell architecture. The optical response of the structures is first evaluated using generalized Mie theory and subsequently compared with FEM and FDTD simulations, ensuring a consistent description of their electromagnetic and thermal behavior. To systematically explore the design space, we define a multi-objective figure of merit that integrates absorption efficiency, absorption cross section, and polymer-layer thickness. We further introduce a thermal gain parameter that quantifies energy densification and complements the analysis of thermal directionality. Our results show a near-optimal structure, capable of producing a temperature rise of 20–23 K, with 67% of the thermal flux directed toward the exposed metallic surface and outperforming spherical nanoparticles by 50% in terms of focusing enhancement. This design preserves geometric simplicity and high symmetry while delivering robust thermal asymmetry, thereby facilitating experimental implementation. Beyond photothermal therapies, the proposed methodology constitutes a versatile platform for the rapid screening and optimization of layered nanostructures, adaptable to diverse materials, excitation wavelengths, and functional objectives in nanophotonics.

**Highlights:**

- Iris-gated core–double–shell nanoparticles enable directional heat transport.
- A multi-objective framework efficiently screens the nanoparticle design space.
- Up to 67% of the thermal power is redirected toward the exposed hemisphere.
- Thermal focusing reaches a 50% gain over the symmetric reference particle.
- Optical absorption varies by less than 5% upon particle reorientation.

## 1. Introduction

Thermoplasmonics has established itself as a versatile platform for generating highly localized temperature increases by exploiting the strong light–matter interaction supported by plasmonic nanostructures[1]. When metallic nanoparticles are excited at their Localized Surface Plasmon Resonances (LSPRs), incident electromagnetic energy is efficiently converted into heat through resistive dissipation[2,3]. A key strength of this mechanism lies in its high degree of tunability, both the spectral position and magnitude of the plasmonic resonance can be engineered by adjusting particle size, morphology, and composition[4]. This tunability has enabled novel applications across nanoscience ranging from photocatalysis and solar energy harvesting to imaging and nanomedicine, among others[5].

In this work, we focus specifically on Photothermal Therapy (PTT), an emerging clinical modality for cancer treatment [6]. In this context, the thermoplasmonic performance of the nanostructures is governed by stringent biological and technological constraints. The primary goal of PTT is the selective ablation of malignant cells via highly localized heating, ideally achieving single-cell or even subcellular resolution. For this purpose, plasmonic tunability is essential. Absorption must be engineered within the biological transparency windows (NIR I-II), where optical attenuation is reduced[7]. Under such conditions, water can reasonably be taken as the surrounding medium and biocompatible dielectric materials, such as Polydimethylsiloxane (PDMS) and polymethyl methacrylate (PMMA), are often employed for structural components[8,9]. This modelling approach, widely adopted in theoretical/numerical studies of plasmonic nanoparticles for PTT[10–13], provides an adequate description of optical and thermal behavior while maintaining the computational efficiency required for extensive parameter screening.

A central challenge in PTT is related to spatial specificity[14]. Although biochemical functionalization governs biological targeting, most homogeneous nanostructures, even those with anisotropic shapes, yield nearly isotropic temperature fields under illumination [1,2,11]. This isotropy can lead to off-target heating and potential damage to the surrounding healthy tissue. To overcome this limitation, hybrid architectures such as Janus particles or composite nanoparticles have been explored. Their material heterogeneity enables the combination of distinct functionalities (chemical, optical, or catalytic) within a single colloid [15]. Such hybrid platforms have been successfully employed in drug delivery, biosensing, and catalysis [16], where multifunctionality provides a clear advantage.

In the context of thermoplasmonics, material heterogeneity introduces an additional degree of freedom: by appropriately combining plasmonic components with materials of contrasting thermal conductivities (high and low), it becomes possible to generate asymmetric temperature profiles under both continuous and pulsed excitation[8,9,17]. Compared to fully symmetric counterparts, such designs can preferentially redistribute heat toward specific spatial regions, enabling directional thermal flux and heat focusing. Nevertheless, the total heat generated by an individual nanoparticle remains limited by its volume, often requiring dense colloidal suspensions for effective therapy. This requirement imposes tight constraints on synthetic reproducibility and monodispersity, particularly for complex Janus-type geometries, thereby motivating the search for simpler architectures capable of producing robust thermal asymmetry.

Here, we propose a minimal yet effective strategy to address this challenge: an Iris-gated Core–Double–Shell (ICDS) nanoparticle. We begin with a highly symmetric stratified sphere, consisting of a glass core to redshift the plasmon resonance into the first biological window (NIR-I), a gold shell serving as the primary heat source, and a polymer overshell acting as thermal insulator. Then, we introduce a single, controlled and minimally sized geometric perturbation: a conical iris aperture in the polymer overshell. This minimal symmetry breaking enables both directional thermal transport and enhanced thermal focusing while preserving fabrication simplicity.

To explore the space of geometrical parameters efficiently, we followed a strategy based on a multi-objective figure of merit that drastically reduces the initial search domain and computational cost. Our analysis reveals that ICDS nanoparticles can simultaneously maximize temperature rise, directional heat flux, and local energy densification, all while preserving high structural symmetry and fabrication feasibility[18–22] . Beyond its implications for PTT, the approach provides a generalizable methodology for the rapid screening of layered nanostructures across different material combinations and excitation wavelengths, thereby establishing a flexible design principle for broader nanophotonics applications.

## 2. Methods

We employed a combination of analytical electrodynamics, full-wave numerical solvers, and finite-element thermal modeling. Specifically, optical simulations were conducted using generalized Mie theory, Finite-Difference Time-Domain (FDTD, Ansys Lumerical), and Finite Element Method (FEM, COMSOL Multiphysics[23]). All analytical calculations were carried out with freely available software packages[24].

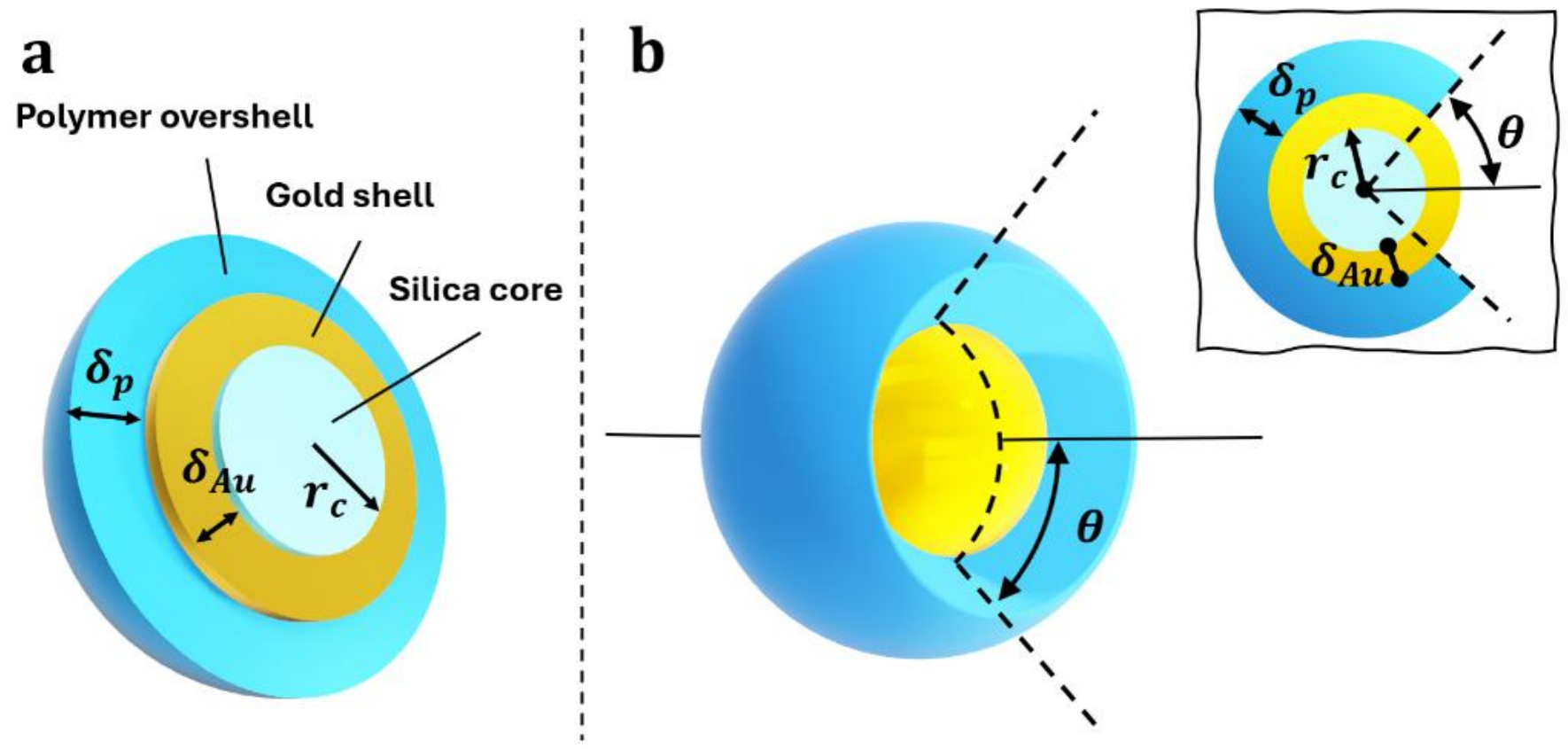

Figure 1. (a) Illustration of the precursor core-double-shell structure defined by the core radius ($r_c$), the gold thickness ($\delta_{Au}$) and the polymer thickness ($\delta_p$). The materials from the core to the overlayer are silica, gold and PDMS respectively. (b) Illustration of the ICDS consisting of a core-double-shell precursor with a conical aperture characterized by the half-opening angle ($\theta$). Upper inset: cross-sectional view of the ICDS.

### 2.1 Optical Modeling of Core-Double-Shell Nanoparticles

Figure 1(a) shows the schematics of the symmetric precursor Core–Double–Shell (CDS) nanoparticles. The structure consists of a silica core ($n = 1.51$), a gold shell and an outer polymer layer. For the purpose of this work, the polymer coating is taken to be PDMS ($n = 1.41$), selected due to its silica-like refractive index and low thermal conductivity. The geometry is defined by the core radius ($r_c$), the gold thickness ($\delta_{Au}$) and the polymer thickness ($\delta_p$), and it is immersed in water ($n = 1.33$). The optical constants of gold were taken from Johnson & Christy data set[25].

To characterize the CDS optical response, we used the generalized Mie theory, which provides an analytical solution for stratified spherical particles. The implementation adopted here, corresponds to the MatScat package developed by Jan Schäfer, freely available through the MATLAB Central File Exchange [24]. The Mie formalism requires a truncation threshold, for the infinite multipolar expansion of the scattering coefficients at a finite order $M$. In the original numerical implementation employed here, this threshold is determined through a convergence implementation-specific parameter, *conv,* according to ref [24] and follows the expression:

$$M = ceil\left(conv\left(x + 4x^{\frac{1}{3}} + 2\right)\right) \tag{1}$$

where $x = 2\pi a/\lambda$ is the size parameter, with $a$ being the particle radius and $\lambda$ the wavelength of the incident light within the surrounding medium. For the particle sizes and wavelength considered here, we set $conv = 1$, yielding truncation orders in the range $M \in [7,10]$ for $x \in [1, 2]$. This means that, up to 10 multipolar orders were considered in our calculations. Convergence was validated by increasing the truncation order beyond $M = 10$ and confirming that no appreciable changes were observed in the calculated optical response (Figure S1).

## 2.2 Electromagnetic Modeling of Iris-gated Core-Double Shell Nanoparticles

Figure 1(b) shows a scheme of the ICDS nanoparticle. As can be seen in the upper inset, the structure maintains the same parameters as the CDS reference (core radius, gold shell thickness, and polymer properties), but incorporates an "eyelid-like" aperture, i.e., a conic opening that defines the extent of the polymeric coverage. This opening is defined by a polar half-angle, $\theta$, corresponding to the spherical cap removed from the polymer coating.

In contrast to CDSs, ICDSs were analyzed using FEM simulations in COMSOL Multiphysics 6.1[23] . The workflow followed two sequential steps: (i) light–matter interaction and (ii) thermal transport. In the electromagnetic stage, Maxwell's equations were solved using COMSOL radiofrequency module. The ICDS geometries were embedded in water, and the surrounding simulation domain was limited by perfectly matched layers to suppress undesired reflections. Illumination was provided by an unpolarized plane wave of intensity I = $0.1\ mW \cdot \mu m^{-2}$, a standard reference value widely employed in thermoplasmonic studies[9,11,26,27]. A free tetrahedral mesh was used in the surrounding medium, with maximum element size constrained to λ/8 to ensure accurate representation of near-field features and geometric curvature. Although values up to λ/5 are often considered adequate for generic plasmonic systems[23], a stricter meshing criterion was selected to ensure accuracy and robustness across all ICDS designs. To resolve the gold shell geometry, a maximum element size of 1 nm was used. To test and validate the FEM results, independent simulations were performed using Lumerical FDTD under a progressively more refined mesh.

### 2.3 Thermal Modelling

Thermal simulations were carried out using the *heat transfer in solids* interface in COMSOL v6.1[23]. The heat rate, Q, computed in the electromagnetic stage was taken as the heat source term in the heat equation.

$$\rho c_p \partial_t T = k\Delta \mathrm{T} + \mathrm{Q} \tag{2}$$

where $\rho$, $c_p$ and $k$ are the mass density, specific heat and thermal conductivity respectively. Thermal boundary resistance at the material interfaces was explicitly accounted for through a finite interfacial thermal conductance, allowing for a temperature discontinuity across the interfaces while preserving heat-flux continuity. In the calculations we considered the gold/PDMS and PDMS/water interfacial thermal conductances, 60 and 50 $\frac{\mathrm{MW}}{\mathrm{m^2K}}$ [28,29]. The gold/water and silica/gold values were 200 and 170 $\frac{\mathrm{MW}}{\mathrm{m^2K}}$ [30,31], respectively. Heat flux boundary conditions were imposed at the system's outer surfaces of the domain, with the transfer coefficient $h = k_{water}/L$, where $L$ is the size of the simulation region. Given the nanoscale dimensions, small fluid volumes, and moderate temperature increases involved (up to a few tens of degrees), heat conduction was treated as the dominant transport mechanism in accordance with established thermoplasmonic literature[1]. Additional details on both the electromagnetic and thermal modelling strategies are provided in Sections I and II in the supplementary document.

## 3. Results and Analysis

### 3.1. Optical study of the precursor Core-Double-Shell structure.

The ICDS structure is defined by four independent structural parameters, resulting in a broad design space whose exhaustive exploration would be computationally expensive. To reduce this complexity while retaining the physical requirements of the target ICDS structures before analysis, we perform a two-step preselection of the geometrical parameter space. The first step identifies suitable core and gold-shell dimensions based exclusively on the electromagnetic response of conventional silica–gold core-shell (CS) nanoparticles (inset in Figure 2), ensuring that the absorption resonance remains within the first biological window (NIR-I, 700–950 nm) while presenting sufficiently high absorption and efficiency.

The second step introduces the polymer overlayer and uses the resulting CDS structures to identify configurations that provide a suitable balance between optical performance, particle size, and the geometrical conditions expected to favour thermal directionality after aperture formation. This approach avoids performing computationally intensive thermal calculations over the full four-dimensional design space, while retaining the physical criteria required to identify promising ICDS geometries.

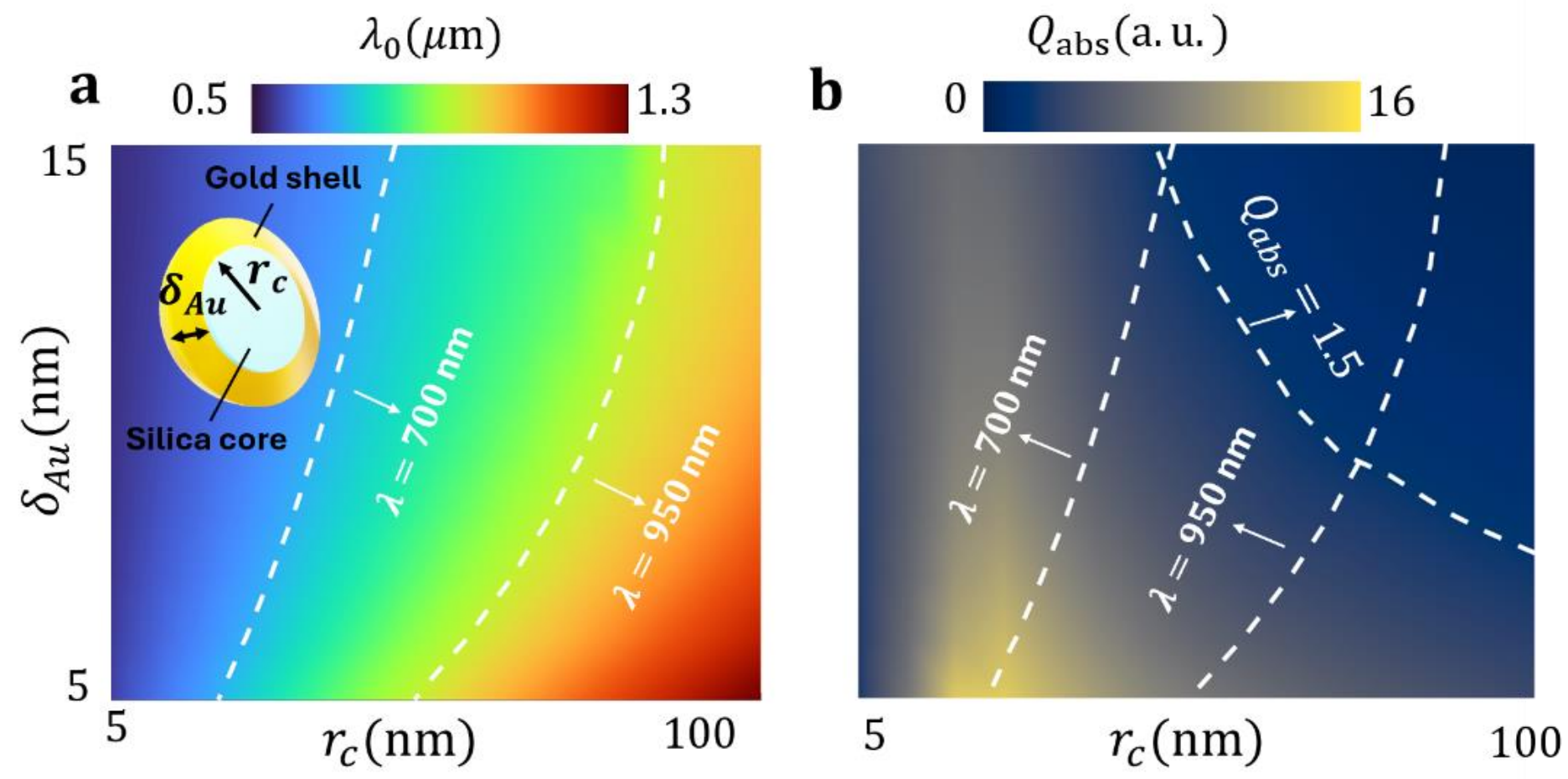

Figure 2. Peak absorption wavelength (a) and absorption efficiency (b) for a silica/gold core-shell as a function of $\delta_{Au}$ and $r_c$. White dashed lines represent the first biowindow band and the limit for an absorption efficiency of 1.5, respectively. Inset: schematics of the CS structure.

We first restricted the design space according to the spectral position of the CS plasmonic resonance, for each combination of $r_c$ and $\delta_{Au}$ as shown in Figure 2a. As expected, the resulting peak wavelengths exhibit a clear dependence on both geometrical parameters. Increasing core radius produces a systematic redshift of the plasmonic response while gold-shell thickness produces a blueshift. Imposing the requirement that the absorption maximum lies within the NIR-I window (700-950 nm) restricts the parameter space to the region limited by the $\lambda = 700$ nm and $\lambda = 950$ nm contours (white dashed lines in Figure 2a).

However, the resonance position alone is not sufficient to identify suitable geometries, since the photothermal performance also depends on the magnitude of the optical absorption. Thus, we evaluated the absorption efficiency, $Q_{\text{abs}}$, over the same parameter space (Figure 2b). The results show that $Q_{\text{abs}}$ generally decreases as either the core radius or the gold-shell thickness increases. This behaviour reflects the reduction in absorption efficiency associated with increasing particle size and/or increasing gold content, despite the fact that these configurations can maintain their resonances within the NIR-I window. This trend is corroborated by Figure S2 which shows the absorption spectra of CS structures with core radii of 5 nm and 80 nm.

To exclude configurations with low absorption efficiency, we imposed $Q_{\text{abs}} = 1.5$ (black dashed line) as a practical lower bound. This value corresponds to an absorption cross section 50% larger than the geometrical cross section, providing a simple criterion to retain structures with sufficiently strong absorption and avoiding unnecessarily large particles. The intersection of this criterion with the NIR-I resonance condition provides an initial restriction of the design space to $r_c \in [10, 75]$ nm and $\delta_{Au} \in [5, 15]$ nm.

These bounds, obtained from the CS analysis, are therefore adopted as the initial parameter range for subsequent CDS study. Since the PDMS overlayer primarily modifies the dielectric environment surrounding the metallic shell, rather than introducing an additional plasmonic material, the CS analysis provides a physically consistent first restriction of the design space.

Within this restricted parameter space, the second preselection step was performed on the CDS structures by introducing the PDMS overlayer. To identify promising precursors without explicitly evaluating the thermal response of every configuration, we defined a merit function combining the normalized absorption cross section $\bar{\sigma}_{abs}$, the absorption efficiency $\bar{Q}_{abs}$ and the relative polymer thickness $\bar{\delta}_p$:

$$\mathcal{F} = \bar{\sigma}_{abs}^{\gamma_1} \bar{Q}_{abs}^{\gamma_2} \bar{\delta}_p^{\gamma_3} \tag{3}$$

The first two terms favour structures combining large absolute absorption, and hence a high potential for heat generation, with efficient absorption relative to their size. The third term favours sufficiently thick polymer coatings, which are expected to promote thermal asymmetry after aperture formation [9], and is therefore used as a proxy for the directional-heating potential of the final ICDS. Since no unique relative weighting can be assigned a priori to these competing criteria, the exponents $\gamma_1$, $\gamma_2$, and $\gamma_3$ were varied within a moderate range (1–1.5), and the recurrence of each configuration as the global optimum was evaluated. This frequency-based analysis identifies geometries that remain favourable under different weightings rather than relying on a single arbitrarily weighted merit function. Further details on the merit-function definition, weighting analysis, and convergence are provided in Section I and Figure S4 of the Supporting Information.

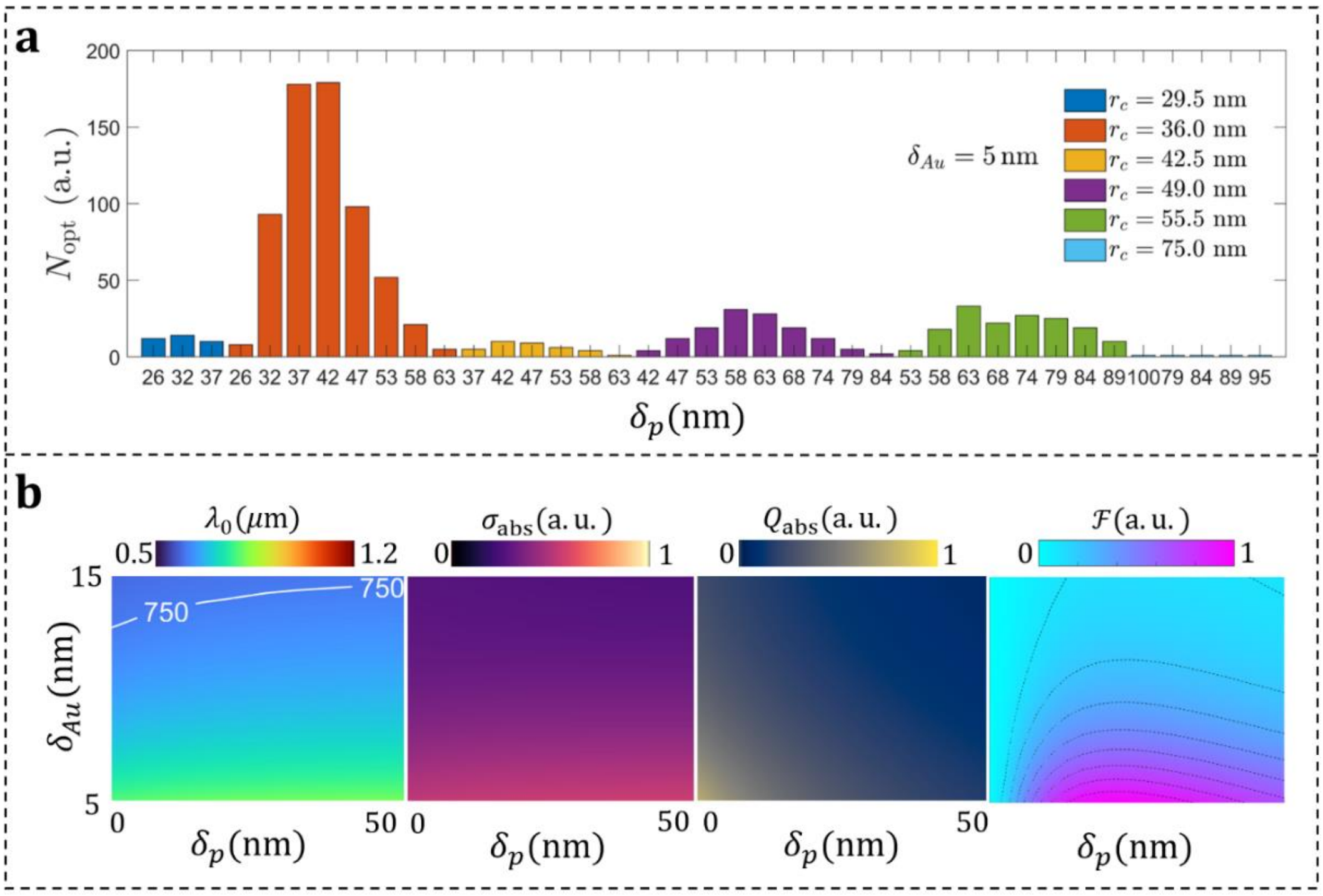

Figure 3. (a) Histogram with the absolute frequency ($N_{opt}$) of optimal configurations for each gamma triplet ($\gamma_1, \gamma_2, \gamma_3$) ranging from 1 to 1.5 as a function of $r_c$ and $\delta_p$. (b) From left to right: peak absorption wavelength, peak absorption cross section, absorption efficiency and trade-off function in terms of $\delta_{Au}$ and $\delta_p$.

The resulting frequency distribution is shown in Figure 3(a), where each bar represents the number of weighting combinations ($N_{opt}$) for which a given CDS geometry was identified as the global optimum of the merit function. The selected configurations are grouped according to their core radius $r_c$, while the horizontal axis indicates the corresponding polymer thickness $\delta_p$.

Remarkably, all configurations emerging as global optima exhibit the minimum gold-shell thickness considered, $\delta_{\mathrm{Au}} = 5\mathrm{nm}$. This systematic selection of thin Au shells is consistent with the optical trends shown in Figure S3, where decreasing $\delta_{\mathrm{Au}}$ generally enhances both the absorption cross section and absorption efficiency over the explored range of core radii and polymer thicknesses. The lower bound of 5 nm was imposed to preserve the validity of the bulk dielectric function used in the Mie calculations, since electron-surface scattering becomes increasingly relevant for thinner metallic layers.

The distribution further shows that the selection is strongly dependent on the core radius. Configurations with intermediate core sizes dominate the frequency analysis, with the highest recurrence obtained for $r_c = 36$ nm and polymer thicknesses concentrated around $\delta_p \approx 37 - 47$ nm. In contrast, configurations with either smaller or larger cores appear considerably less frequently. This behaviour reflects the competition embedded in the merit function: small structures are favoured by their high absorption efficiency but provide lower absolute absorption and limited polymer volume, whereas increasing the particle dimensions enhances the absolute absorption and accommodates thicker polymer layers at the expense of absorption efficiency. Consequently, neither extreme remains optimal when the relative weighting of the three criteria is modified.

To rationalize the origin of this preferential region, Figure 3(b) shows the optical response and merit function for the most recurrent core radius, $r_c = 36\mathrm{nm}$, as a function of $\delta_{\mathrm{Au}}$and $\delta_p$. The resonance map shows that the spectral position is governed predominantly by the gold-shell thickness, whereas variations in the polymer coating produce comparatively smaller shifts. Most of the relevant parameter space remains within the NIR-I window, as indicated by the 750 and 950 nm contours. A similar dependence is observed for $\sigma_{\mathrm{abs}}$, which is substantially more sensitive to the Au-shell thickness than to the polymer coating, reaching its largest values for thin gold shells. This trend is preserved over the complete range of core radii, as shown in Figure S3. The absorption efficiency introduces a stronger size dependence: $Q_{\mathrm{abs}}$ is maximized for thin Au shells and decreases as the overall particle dimensions increase, thereby penalizing configurations with thick polymer coatings and/or large metallic shells.

These competing tendencies are combined in the merit-function map. While the optical terms favor thin Au shells and comparatively compact structures, the polymer-thickness contribution increasingly favors thicker coatings because of their expected potential to enhance thermal asymmetry after aperture formation. Their competition produces a broad maximum at intermediate $\delta_p$ , consistent with the preferential polymer-thickness range identified independently by the frequency analysis in Figure 3(a).

Therefore, the frequency analysis identifies CDS precursors combining thin Au shells, intermediate core radii, and moderate polymer thicknesses as the most robust compromise between absolute absorption, absorption efficiency, and the geometrical potential for thermal directionality. These configurations define the reduced design space considered in the subsequent ICDS analysis, where the asymmetric aperture is introduced and the resulting thermal response is explicitly evaluated.

### 3.2. Photothermal study of the Iris-gated Core-Double-Shell nanoparticle.

Having restricted the CDS design space, we next introduced the conical aperture to evaluate the photothermal response of the asymmetric ICDS structures. Rather than exploring the complete parameter space again, we selected $r_c = 36$ nm and $\delta_{\mathrm{Au}} = 5$ nm, corresponding to the smallest core radius exhibiting the highest recurrence in the frequency analysis and the consistently preferred Au-shell thickness, respectively. Based on the same analysis, polymer thicknesses up to 60 nm were considered, including $\delta_p = 0$ nm as the limiting CS configuration. As illustrated in Figure 4(a), the aperture is defined by the half-opening angle $\theta$, which was progressively varied to control the fraction of the metallic surface exposed to the surrounding medium. The limiting cases $\theta = 0°$ and $180°$ correspond to the CDS and CS configurations, respectively, thereby providing direct reference limits for the intermediate ICDS geometries.

Figure 4(b) summarizes the resulting resonance wavelength, peak absorption cross section, and steady-state temperature increase as functions of $\delta_p$ and $\theta$. Increasing the aperture progressively blueshifts the resonance toward that of the conventional CS particle, as the fraction of the Au surface surrounded by PDMS decreases. Conversely, thicker and more complete polymer coatings produce a redshift due to the higher refractive index of PDMS relative to water. The absorption cross section follows a similar trend, with the largest values obtained for thicker coatings and smaller apertures.

These optical variations translate directly into the thermal response. The largest temperature increases occur for thick PDMS coatings and small apertures, resulting from the combined enhancement of optical absorption and the lower thermal conductivity of PDMS compared with water, which increases thermal confinement around the metallic heat source[3]. Nevertheless, substantial heating is retained over a broad range of opening angles, demonstrating that relevant structural asymmetry can be introduced without suppressing a strong photothermal response.

However, for practical applications, the photothermal response should also remain robust against particle orientation. This requirement is particularly relevant for colloidal systems, where rotational alignment cannot generally be controlled. Therefore, we evaluated the absorption response under particle rotation, as schematically illustrated in Figure 4(c). For each ICDS geometry, the particle was rotated by an angle $\alpha$ relative to the incident unpolarized wave along the z-axis while maintaining the excitation wavelength corresponding to its maximum absorption in the reference orientation (opening fully aligned towards the positive z-axis). Rotational robustness was quantified through an absorption-stability parameter, $S_\alpha$, defined from the maximum relative variation of $\sigma_{\mathrm{abs}}$ over the sampled orientations (see Supporting Information, Section I, for its definition). By construction, $S_\alpha = 100\%$ corresponds to a completely orientation-independent absorption response.

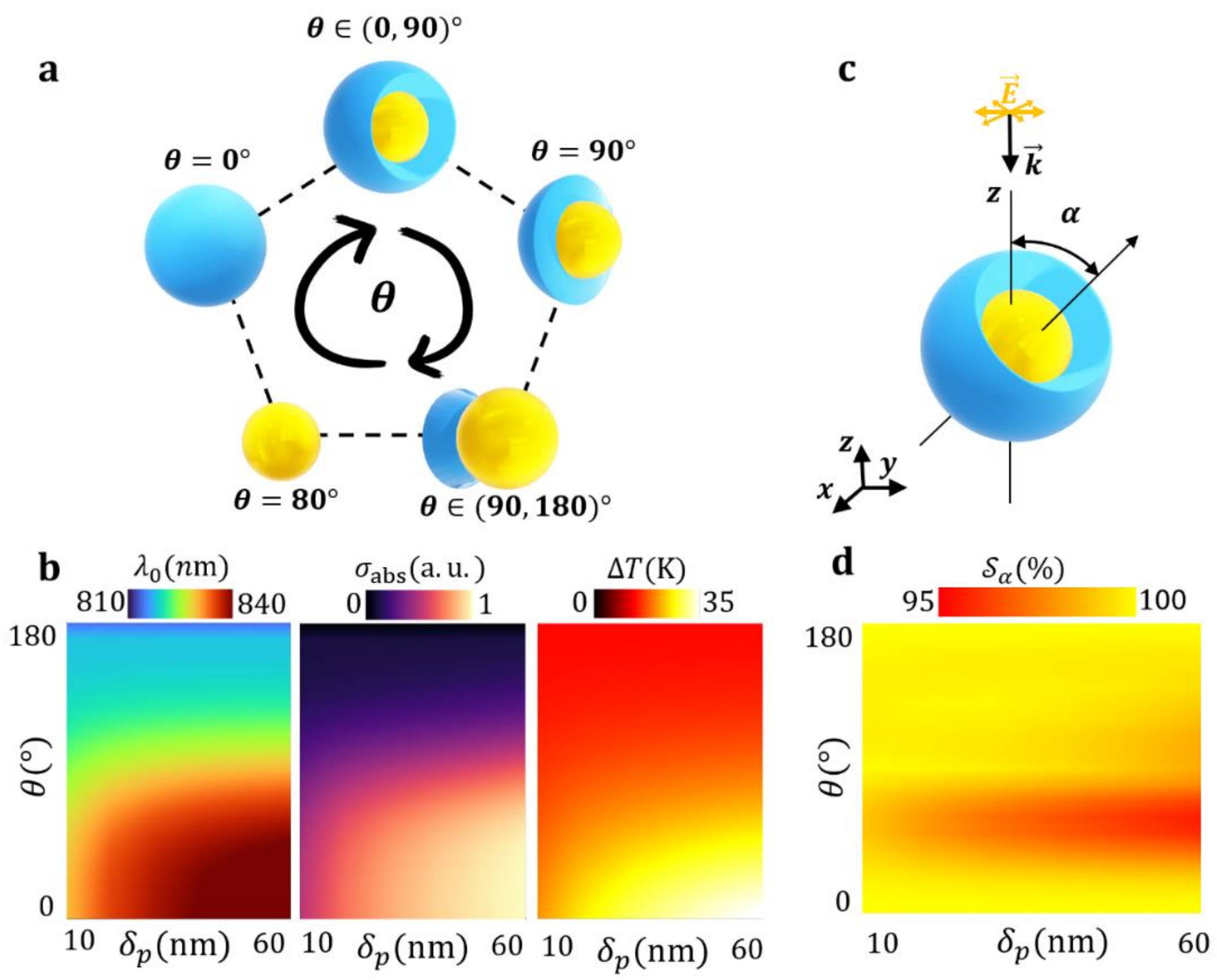

Figure 4. a) Illustration of the ICDS and the procedure followed in the analysis. $\theta$ is the semi-angle, measured from the centre of the particle, that defines the iris. b) Resonance wavelength, absorption cross section and temperature increase maps as a function of the aperture angle $\theta$ and the polymer thickness $\delta_p$. c) Schematic representation of the unpolarized plane wave illuminating the ICDS at an angle $\alpha$, defined as the angle between the incident wavevector $\vec{k}$ and the $z$-axis. d) Stability map as a function of the aperture angle $\theta$ and the polymer thickness $\delta_p$.

Remarkably, the resulting stability map shown in Figure 4(d) remains within $S_\alpha = 95$–$100\%$ throughout the entire investigated $(\delta_p, \theta)$ space. Thus, despite the pronounced geometrical asymmetry introduced by the aperture, particle reorientation modifies the peak absorption by at most approximately 5%. The optical response of the ICDS is therefore essentially insensitive to orientation over the relevant design space. This rotational robustness constitutes an important advantage of the ICDS architecture: thermal asymmetry can be introduced through selective modification of the passive polymer coating while retaining the nearly orientation-independent optical response of the underlying spherical plasmonic core. This contrasts with strongly anisotropic plasmonic geometries, such as nanorods or nanodisks, whose absorption even though strong[4,11] can depend substantially on their orientation relative to the incident field [11,32].

Together with the screening results of Figure 3, this analysis restricts the most promising ICDS designs to a core radius of $r_c = 36$ nm, a gold-shell thickness of $\delta_{Au} = 5$ nm, polymer thicknesses in the range $\delta_p = 30$–$50$ nm, and partial aperture openings within $\theta = 0^o$–$90^\circ$. Within this parameter space, the structures preserve NIR-I absorption, substantial photothermal heating, and high rotational robustness. These geometries are therefore retained for the subsequent thermal analysis, where the magnitude and spatial directionality of the generated temperature field are explicitly quantified.

### 3.3. Analysis of the thermal flux in the stationary regime.

Once a favourable range of ICDS geometries is identified, we quantify how the asymmetric polymer coating redistributes the conductive heat flux around the nanoparticle. To this end, we introduce two complementary metrics: the directional power fraction, $\boldsymbol{\eta}$, which quantifies the fraction of the total dissipated thermal power directed within a prescribed solid angle, and the thermal focusing gain, $\mathbf{G}$, which measures the corresponding increase in power density relative to the fully symmetric reference particle. Their complete mathematical derivation and numerical implementation are provided in Sections II and III of the Supporting Information.

As illustrated in Figure 5(a), the ICDS is enclosed by a concentric spherical control surface located at a fixed radial distance $\boldsymbol{\Delta} = \mathbf{15}$ nm from the particle, as it is a representative separation for antibody-functionalized nanoparticles interacting with cell membranes[33]. A sensitivity analysis is provided in Section II of the Supporting Information (Figure S7), confirming the robustness of the conclusions with respect to the choice of $\boldsymbol{\Delta}$. The outward conductive heat flux is integrated over a spherical cap defined by the polar half-angle $\boldsymbol{\beta}$. The directional power fraction is then defined as

$$\eta(\beta) = \frac{\iint_{\Omega(\beta)} \vec{\Phi} \cdot \vec{n}\, dS}{\iint_{\mathbb{S}^2} \vec{\Phi} \cdot \vec{n}\, dS} \tag{4}$$

where $\theta$ defines the ICDS aperture, $\vec{\Phi}$ is the conductive heat-flux vector, and $\mathbb{S}^2$ is the complete control surface. Thus, $\eta$ represents the fraction of the total outward thermal power contained within the solid angle subtended by $\beta$.

Figure 5(b) shows the calculated $\eta(\beta, \theta)$ for the ICDS geometry. The black dashed contours identify the values of $\beta$ required to collect 25%, 50%, and 75% of the total thermal power, whereas the corresponding values for the symmetric reference particle are indicated by the red dashed lines. For the asymmetric ICDS, these contours are displaced toward smaller $\beta$, showing that a given fraction of the total thermal power crosses a smaller portion of the control surface than in the symmetric case. Consequently, the aperture redistributes the conductive heat flux toward the exposed hemisphere. As shown in Figure 5(b), the largest deviations from the symmetric case occur for ICDS structures with an iris in the range $\theta = [60, 80]^0$. For these geometries, the fraction of power directed toward the upper hemisphere ($\beta = 90^0$) increases significantly, revealing a pronounced thermal redirection effect.

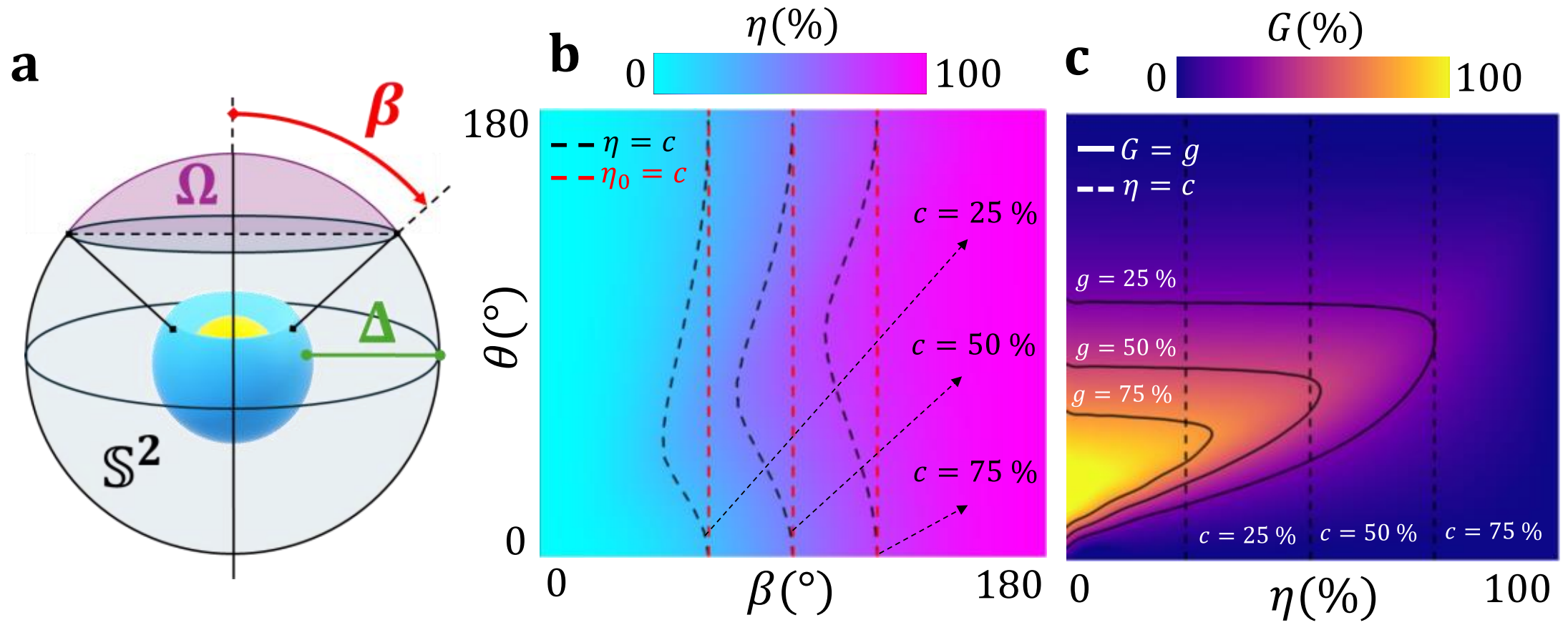

Figure 5. a) Geometrical construction employed to quantify the directional response of the ICDS nanoparticle. The total outward power is collected over the closed integration control surface $\mathbb{S}^2$. A spherical cap of semi-aperture $\beta$ (purple domain with solid angle $\Omega$) is used to evaluate the power fraction $\eta$. The control surface $\mathbb{S}^2$is located at a distance $\Delta = 15$ nm from the ICDS surface. b) Map of $\eta(\beta, \theta)$ expressed as a percentage. Black dashed curves trace the regions where the power fraction is constant with $\eta = 25, 50, 75$ %, while red dashed curves indicate the corresponding values for the symmetric case. c) Gain map $G(\eta, \theta)$. Solid black isolines mark the regions of constant gain for 25%, 50%, and 75%; vertical dashed lines highlight the regions along which the power fraction $\eta$ is constant with the same percentages.

To quantify this spatial concentration, we introduce the thermal focusing gain, $G(\eta, \theta)$, defined as the percentage increase in the average surface power density required to transmit a given power fraction $\eta$, relative to the symmetric reference particle. Thus, $G = 50\%$, for example, indicates that the same fraction of the total thermal power is transmitted with an average surface power density 1.5 times larger than in the symmetric case. The complete derivation is provided in Section III of the Supporting Information. In particular, $\eta = 50\%$ provides a convenient reference because, for the symmetric particle, half of the total power crosses each hemisphere. Deviations from this condition therefore provide a direct measure of the heat-flux redistribution induced by the aperture. While $\eta$ quantifies directional redistribution, it does not directly measure the spatial concentration of the transported power.

Figure 5(c) shows the resulting gain map obtained from the directional power distributions in Figure 5(b). Positive values of $G$ indicate that the same fraction of the total thermal power is transmitted through a smaller surface area than in the symmetric reference, and therefore with a larger average surface power density. A pronounced region of enhanced focusing emerges at intermediate opening angles, reflecting the interplay between the extent of the insulating polymer coverage and the resulting redistribution of the conductive heat flux. For the reference condition $\eta = 50\%$, the maximum focusing gain reaches approximately $G = 50\%$, corresponding to an average surface power density 1.5 times larger than that of the symmetric particle.

From these metrics, we identify an intermediate range of aperture angles as the most favourable regime. Within this region, up to approximately 67% of the total thermal power is directed toward the exposed hemisphere ($\beta = 90°$), while the focusing gain reaches approximately 35% when evaluated over the same hemisphere. Moreover, when the comparison is performed at a reference power fraction of $\eta = 50\%$, the maximum gain approaches 50%. Thus, the ICDS simultaneously redirects a larger fraction of the generated thermal power toward the exposed side and concentrates this power over a smaller effective area than the symmetric reference.

Taking all metrics into account, including temperature rise, heat-flux directionality, thermal focusing, rotational stability, and optical performance, ICDS structures with aperture angles in the range $\theta = 60° - 80°$ provide the most balanced overall response. These configurations achieve temperature increases of approximately 20–23 K, direct up to 67% of the total thermal power toward the exposed hemisphere, and increase the average surface power density over this hemisphere by approximately 35% relative to the symmetric reference. Moreover, when considering a fixed power fraction of $\eta = 50\%$, the focusing gain reaches approximately 50%, corresponding to an average surface power density 1.5 times that of the symmetric particle. These results restrict the most favourable design space to $r_c = 36$ nm, $\delta_{\mathrm{Au}} = 5$ nm, $\delta_p = 30 - 50$nm, and $\theta = 60° - 80°$, providing a balanced compromise between optical performance, photothermal heating, thermal directionality, and rotational robustness.

### 3.4. Thermal directionality under pulsed illumination.

Finally, we investigate the ICDS behavior under pulsed excitation. Pulsed thermoplasmonics is particularly relevant in heterogeneous biological environments, where differences in thermal relaxation times can be exploited through pulse engineering to control the spatial and temporal extent of heating [34].

The characteristic thermal response time of the selected ICDS was first determined from its transient response under CW illumination. Fitting the temperature rise to an error-function profile [1] yields a characteristic heating time of approximately 30 ns (Figure S8), placing the thermal dynamics of the structure in the nanosecond regime. We therefore consider this value as the pulse duration, allowing the transient directional response of the particle to be resolved before thermal diffusion progressively drives the system toward equilibrium. The pulse fluence was set to 0.3 mJ $cm^{-2}$, corresponding to a time-averaged power density of 0.1 mW $\mu m^{-2}$, consistent with the reference irradiation conditions considered above[8].

Figure 6(a) shows the temporal evolution of the absorbed power and temperature under pulsed excitation. Temperatures were monitored at the maximum-temperature position within the Au shell, $T_{\mathrm{Au}}$, and at two points located along the symmetry axis, 15 nm from the exposed and polymer-covered sides of the particle, $T_+$ and $T_-$, respectively. This distance provides a representative length scale for the separation between a biofunctionalized nanoparticle and a nearby biological target[35]. The thermal response is delayed with respect to the absorbed optical power, reflecting the finite thermal diffusion timescale of the system [10]. Moreover, the three temperature traces exhibit distinct temporal responses: heat propagates more rapidly toward the exposed aqueous region, whereas transport through the PDMS-covered side is slowed by the substantially lower thermal conductivity of the polymer. Consequently, temperature differences exceeding 13 K develop across the polymer-coated region, while the front–rear temperature contrast reaches approximately 8 K, demonstrating that the directional thermal response observed under CW illumination is retained during pulsed excitation.

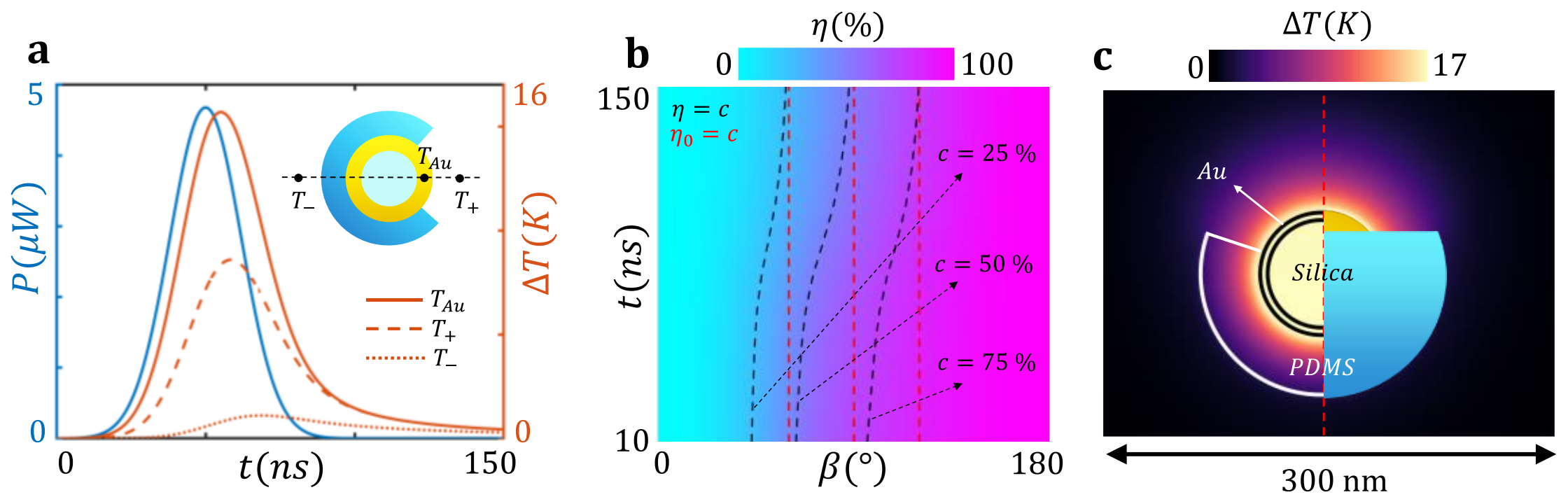

Figure 6. a) Time evolution of the total power dissipated in the gold shell (blue curve, left axis) and the corresponding temperature increases at three key locations: in the gold shell itself ($T_{Au}$), in the exposed direction at 15 nm from the Au surface ($T_+$), and in the shielded region over the control surface ($T_-$). The inset illustrates the geometric location of these points over the ICDS nanoparticle. b) Power fraction $\eta(\beta, t)$. Black dashed curves indicate the condition $\eta = c$, while red dashed curves correspond to $\eta_0 = c$, for coverage fractions $c = 25\%$, 50%, and 75%. c) Temperature map at the instant when the gold shell reaches its maximum temperature. The left half displays the computed temperature field, while the right half is overlaid with a schematic of the nanoparticle cross-section, showing the internal layers (Au and PDMS) and delimiting the physically inaccessible domains.

The temporal evolution of this directionality is quantified in Figure 6(b) using the same directional power fraction introduced for the stationary analysis, now evaluated from the instantaneous conductive heat flux. The dashed curves indicate the integration angles $\beta$required to collect fixed fractions of the total outward thermal power, while the corresponding symmetric values are shown for reference. During the initial heating stage, the ICDS curves are displaced toward smaller $\beta$, demonstrating that a given fraction of the thermal power is confined to a smaller solid angle than in the symmetric case. This anisotropy progressively decreases during thermal relaxation, with the equi-power contours approaching the symmetric reference as heat diffusion homogenizes the temperature field after the optical excitation decays. Thus, under pulsed excitation, thermal directionality is intrinsically time dependent and is strongest during the heating stage.

Figure 6(c) shows the temperature distribution at the instant of maximum Au-shell temperature ($t \approx 53$ ns). A pronounced temperature maximum develops adjacent to the exposed aperture, producing an extended heated region directly accessible from the surrounding medium. In contrast, the PDMS overlayer suppresses heat propagation toward the covered side, maintaining a substantially lower temperature in this region. The transient temperature field therefore confirms the behaviour inferred from the heat-flux analysis: the asymmetric polymer coating acts simultaneously as a thermal barrier and as a means of redirecting heat toward the exposed side of the nanoparticle.

The ICDS retains its directional thermal response under nanosecond pulsed excitation, while the magnitude of the asymmetry evolves according to the competition between optical heating and thermal relaxation. This behaviour, together with the strong NIR-I absorption, rotational robustness, and directional response demonstrated under CW illumination, highlights the potential of the ICDS architecture for spatially and temporally controlled photothermal heating.

## 4. Conclusions

In this work, we have presented a computational methodology and a multi-objective figure of merit tailored to simultaneously evaluate the optical and thermal performance of multilayered nanoparticles. Our proposed framework combines generalized Mie theory with full-wave numerical simulations, enabling a systematic and efficient exploration of the multidimensional design space. Importantly, both the figure of merit and the computational workflow are formulated in a generalizable manner, making them readily extensible beyond the specific case of CDS nanoparticles to a broad range of plasmonic and hybrid architectures.

We also defined a thermal gain parameter, a metric that quantifies the degree of thermal intensification and complements the conventional interpretation of thermal directionality. By establishing a direct connection between directional heat flux and surface power density, this parameter provides a more rigorous description of anisotropic thermal transport in nanoscale systems. This link between directionality and densification of thermal energy, offers a robust physical criterion for evaluating and comparing candidate structures that optimize directional heating.

From the extensive parameter space explored, we identified an optimal ICDS design defined by $(r_c, \delta_{Au}, \delta_p, \theta) = (36\,nm, 5\,nm, 40\,nm, 70°)$ . This structure delivers a maximum temperature increase of 20–23 K, a thermal directionality of up to 67% of the total power flux toward the upper hemisphere, and a focusing gain of approximately 50%, corresponding to a 1.5-fold increase in surface power density compared to the symmetric case. This performance is achieved while maintaining high geometric simplicity and symmetry, with the absorption remaining largely insensitive to particle orientation. Furthermore, this architecture is feasible from an experimental standpoint as supported by previous works[18–22] . The combination of strong thermal asymmetry, directional focusing, optical robustness, and structural feasibility makes the ICDS design a candidate for experimental testing as a directional thermoplasmonic design for photothermal therapy applications.

Furthermore, the rotational stability remains consistently high across the explored aperture range, with $S_\alpha$values above 95% and only a weak dependence on $\theta$ (Figure 4d). This confirms that the thermal asymmetry introduced through the polymer aperture can be achieved without significantly compromising the orientation-independent optical response of the particle.

Finally, the rapid screening and analysis methodology developed in this work is expected to be of interest to both theorists and experimentalists working on the development of advanced specialized nanostructures. Beyond the specific case of plasmonic photothermal therapy, the general framework can be readily adapted to other applications involving different operating wavelengths, materials, and performance constraints, including sensing, photocatalysis, energy harvesting, and nanoscale thermal management. As such, the approach presented here provides a versatile strategy for guiding design and optimization of next-generation thermoplasmonic technologies.

**Author contributions**

Javier González-Colsa: Conceptualization, Methodology, Software, Formal analysis, Investigation, Data curation, Visualization, Writing - original draft, Writing - review and editing.

Fernando Bresme: Conceptualization, Methodology, Supervision, Writing - review and editing.

Pablo Albella: Conceptualization, Methodology, Supervision, Funding acquisition, Writing - review and editing.

**Data availability**

The datasets and MATLAB scripts supporting the findings of this study are publicly available in Zenodo at https://doi.org/10.5281/zenodo.23014870. Additional data not included in the repository are available from the corresponding author upon reasonable request.

**Declaration of generative AI and AI-assisted technologies in the manuscript preparation process**

During the preparation of this work, the authors used ChatGPT (OpenAI) to assist with language review and organization. After using this tool, the authors reviewed and edited the content as needed and take full responsibility for the content of the published article.

**Acknowledgements**

P.A. acknowledges support by the Spanish Ministerio de Ciencia e Innovación under project MOPHOSYS (grant no. PID2022-139560NB-I00). J. G.-C. acknowledges J. Cachón and M. P. Ogáyar for the scientific discussions on heat transport in biological tissues and financial support from the Xunta de Galicia through the Programa de axudas de apoio á etapa de formación posdoutoral (ED481B-2026-093), funded by the Consellería de Educación, Ciencia, Universidades e Formación Profesional. FB thanks the UK Leverhulme Trust (Grant No. RPG-2018-384).

# Supporting Information

# Design Principles for Tailoring Heat Transport via Iris-Gated Core–Double–Shell Nanoparticles in the Context of Photothermal Therapies

*Javier González-Colsa[a,1,]*, Fernando Bresme[b], Pablo Albella[a,]**

E-mail: javier.gonzalez.colsa@uvigo.gal

E-mail: pablo.albella@unican.es

[a]Group of Optics, Department of Applied Physics, University of Cantabria, 39005 Santander, Spain.

[b]Department of Chemistry, Molecular Sciences Research Hub, Imperial College London, London W12 0BZ, UK.

Present address: [1]CINBIO, University of Vigo, Campus Universitario de Vigo, Lagoas Marcosede, Vigo 36310, Spain.

## I.-Electromagnetic analysis of Core-Double-Shell structures.

To assess the convergence of the multipolar expansion, we considered one of the largest core-shell particles analyzed ($r_{\mathrm{core}} = 100$ nm, $\delta_{\mathrm{Au}} = 15$ nm), which represents one of the most demanding cases in terms of size parameter. As can be seen in Figure S1a, the absorption cross-section spectra obtained for $conv = 0.8$, 1.0, and 1.2, corresponding to truncation orders $M =$ 8, 10, and 12, respectively, are essentially indistinguishable. In addition, the integrated relative spectral error shown in Figure S1b, evaluated with respect to the $conv = 1.5$ reference solution, rapidly decreases with increasing $conv$, confirming that $conv = 1$ provides a converged optical response.

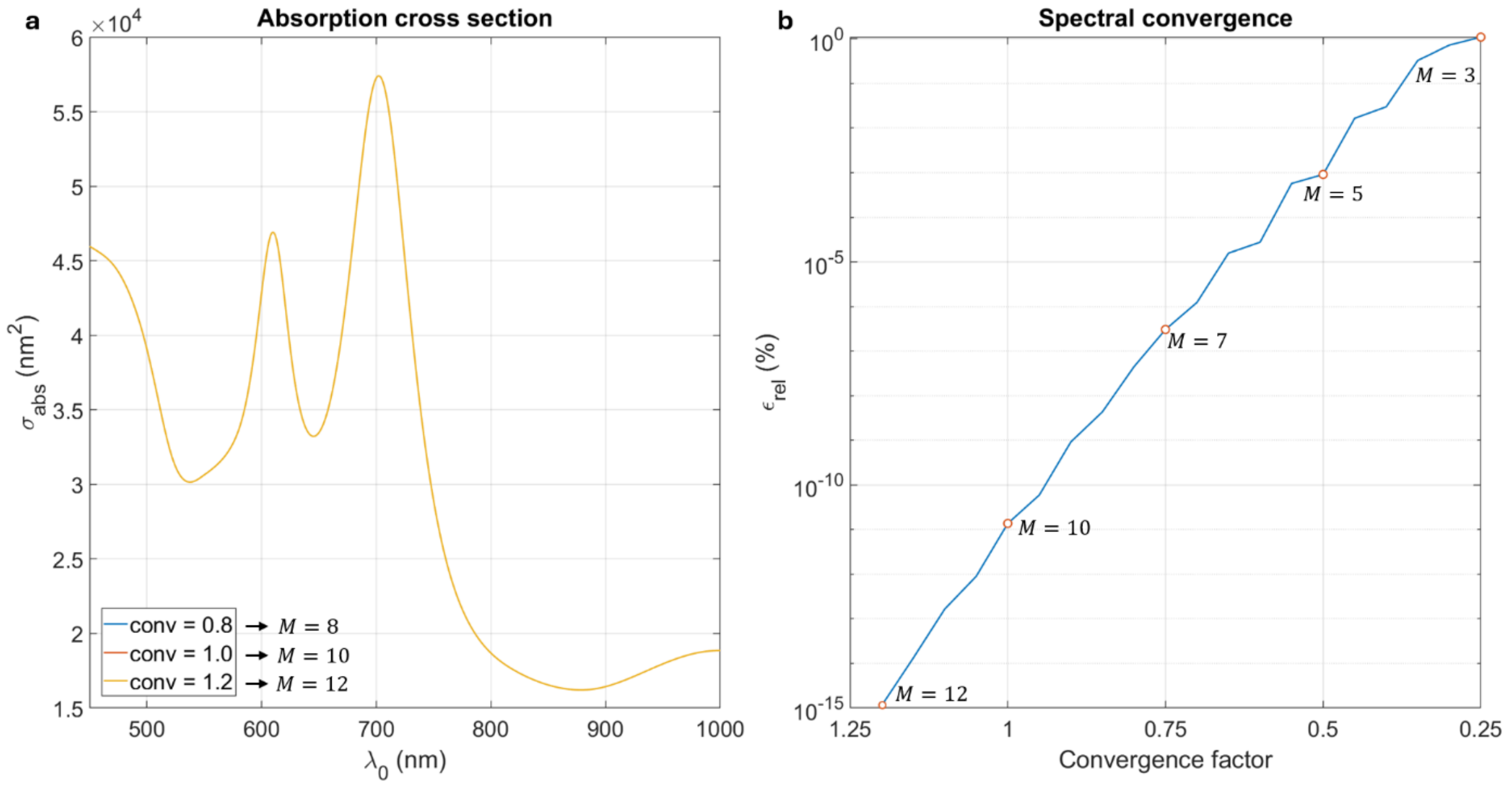

Figure S1. Convergence analysis of the multipolar expansion used in the Mie calculations. (a) Absorption cross-section spectra of the silica/gold core-shell nanoparticle in water for three representative convergence factors, $conv = 0.8$, 1.0, and 1.2, corresponding to truncation orders $M = 8$, 10, and 12, respectively. (b) Relative integrated spectral error, $\epsilon_{\mathrm{rel}}$, calculated with respect to the reference solution obtained with $conv = 1.5$, as a function of the convergence factor. Selected points indicate the corresponding multipolar truncation orders $M$.

To guide the selection of the structural parameters for the more complex core–shell–shell geometry, we begin by analysing a simpler system: the traditional core–shell nanoparticle. This simplification is justified by the fact that the overshell in the definitive design is composed of a polymeric material, which is expected to introduce only minor deviations in the absorption cross-section compared to the standard core–shell configuration. This is because the main absorption mechanism is predominantly governed by a plasmonic mode localized in the gold shell. The design criterion employed is to target structures that resonate within the first biological transparency window (approximately 750–950 nm), which is optimal for biomedical applications such as photothermal therapy. This is one of the motivations for focusing on core–shell structures, in addition to the fact that they admit fully analytical electromagnetic solutions, which facilitates physical description while maintaining low computational costs.

Figure S2 illustrates the absorption cross-sections for a set of limiting cases. In panel (a), we show results for a core–shell nanoparticle with a fixed core diameter of 10 nm and varying gold shell thicknesses. As can be observed, the spectral response resembles that of a solid gold sphere, with thicker shells yielding higher absorption values. However, the overall absorption remains weak, and the resonance peaks lie outside the first therapeutic window. Based on these results, we conclude that configurations with small core sizes are suboptimal for our application, and we therefore focus on core–shell structures with larger core radii in subsequent analyses.

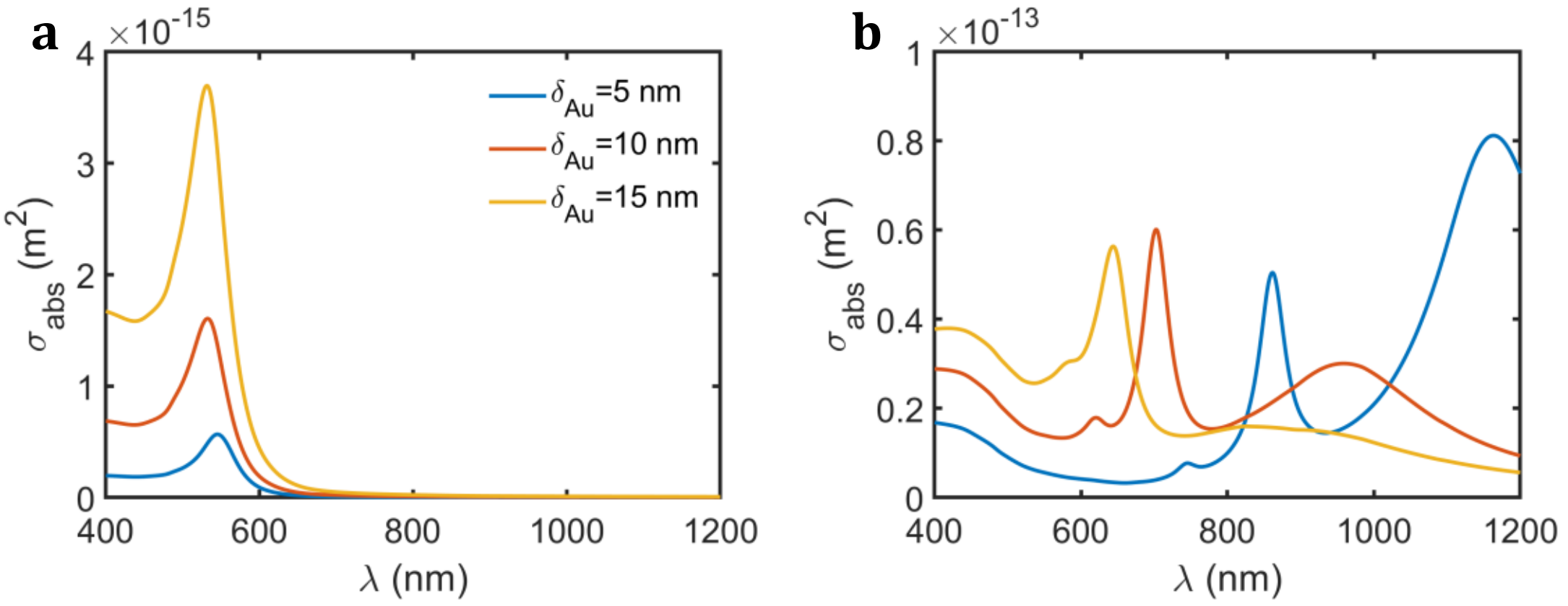

Figure S2. Spectral absorption cross sections for $\delta_{Au} = 5, 10, 15$ nm in the case of $r_c = 5$ nm (a) and $r_c = 80$ nm (b).

In contrast, panel (b) of Figure S2 presents the absorption cross-section for a core–shell nanoparticle with a large core diameter of 160 nm. As shown, when the gold shell is only 5 nm thick, the plasmonic resonance lies far beyond the second biological transparency window (typically defined between 1000–1400 nm). Although the resonance shifts back toward the therapeutic window for thicker gold coatings, its magnitude is significantly lower. Even though such large structures can exhibit strong absolute absorption, their absorption efficiency, defined as the absorption cross-section normalized by the geometrical cross-section, is expected to be low. In some cases, it may even fall below the efficiency of smaller structures due to the relatively larger volume that does not contribute effectively to plasmonic absorption. For this reason, core diameters larger than 160 nm are determined suboptimal and will not be considered further in the design set.

About the gold shell thickness, we define 5 nm and 15 nm as practical lower and upper bounds, respectively. Shells thinner than 5 nm require semiclassical corrections to the dielectric function due to surface electron scattering effects[1], which complicate both modelling and interpretation. On the other hand, gold shells thicker than 15 nm tend to induce a spectral blueshift in the plasmonic response, typically accompanied by a reduction in the peak absorption cross-section [2].

Given our objective, to maximize photothermal conversion while balancing absorption efficiency, these limits represent a reasonable and physically motivated design range. It is worth noting, however, that slight deviations from these assumptions are unlikely to produce significant changes in the overall conclusions.

Once the approximate design limits for the underlying structure (i.e., the traditional core–shell nanoparticle) have been established, we proceed to perform a systematic study based on Mie theory for a double-layered stratified sphere, commonly referred to as a Core–Double–Shell (CDS) nanoparticle. In this analysis, we consider configurations where the core radius, $r_c$, ranges from 10 to 75 nm, and the gold shell thickness, $\delta_{Au}$, spans from 5 to 15 nm. Regarding the overshell, its thickness, $\delta_p$, is varied between 0 and 100 nm. This parameter space is designed to explore the combined influence of geometric and material features on the optical and potential thermal behaviour of the CDS structures, while remaining within the practical bounds previously identified.

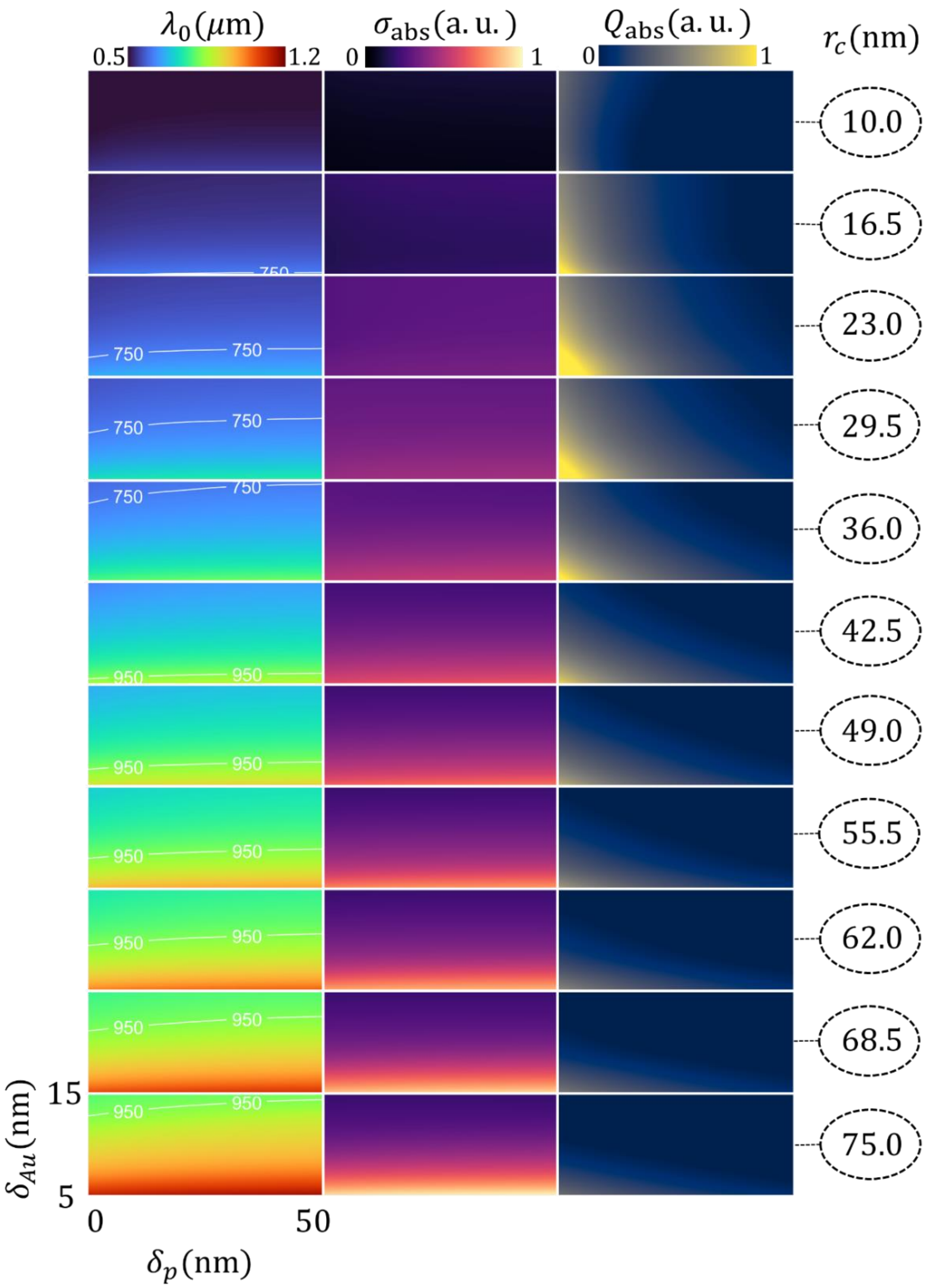

Figure S3. Colour maps showing the spectral and absorption characteristics of core–shell nanoparticles as a function of the gold shell thickness $\delta_{Au}$and the polymer coating thickness $\delta_p$, for a series of core radii ($r_c$) indicated on the right-hand side. Each row corresponds to a specific core radius, increasing from top to bottom. The three columns respectively represent: (left) the resonance peak wavelength $\lambda_0$ in micrometers, (centre) the normalized absorption cross-section in arbitrary units $\bar{\sigma}_{abs}$, and (right) the absorption efficiency $\bar{Q}_{abs}$, also normalized. White contour lines overlaid on the $\lambda_0$ maps indicate the iso-wavelength levels at 750 nm and 950 nm, which serve as references for identifying plasmonic regimes of interest in the visible and near-infrared ranges.

Figure S3 provides a detailed visualization of the optical response of CDS nanoparticles, considering variations in the core radius, gold shell thickness, and polymer coating thickness. Due to the very limited variation observed at high polymer thicknesses, only the region corresponding to $\delta_p \in [0{,}50]$ nm is shown. Including larger values of the polymer shell would unnecessarily reduce contrast across the maps without offering significant additional insight, as the qualitative behaviour of the system remains essentially unchanged.

Each row corresponds to a fixed value of $r_c$, increasing from top to bottom from 10 to 75 nm, while each column represents a different quantity of interest: the resonance peak wavelength $\lambda_0$(left), the normalized absorption cross-section $\bar{\sigma}_{abs}$ (centre), and the absorption efficiency $\bar{Q}_{abs}$ (right). The data shown are the result of rigorous calculations based on Mie theory for multi-layered spheres[3].

A clear and progressive redshift of the plasmonic resonance is observed as the core radius increases. For small cores ($r_c \approx 10$ nm), the resonance remains in the visible range, far from the biologically relevant near-infrared window. As core radius increases, the peak wavelength shifts toward longer values, entering the first biological transparency window (roughly 750–950 nm). This trend is particularly relevant for $r_c \geq 36$ nm, where the resonance remains confined within this spectral region for a broad range of gold and polymer thicknesses. The white contour lines overlaid on the $\lambda_0$ maps, at 750 nm and 950 nm, serve as visual guides to identify the configurations falling within this therapeutically relevant window.

The behaviour of the normalized absorption cross-section reveals that the overall strength of absorption increases with both core size and gold shell thickness. However, the dependence on the polymer layer is comparatively weak, as anticipated, with only a modest modulation of the absorption across the explored range.

In terms of absorption efficiency, which accounts for the geometrical size of the nanoparticle, the results point to a more complex scenario. Although the absolute absorption increases with particle size, the efficiency is highest for intermediate-sized particles, typically in the range of $r_c \approx 23 - 36$ nm, especially when combined with a gold shell of thinner thickness (around 5-10 nm). This behaviour reflects the competing roles of resonant absorption enhancement and geometrical reduction: large particles collect more energy in absolute terms but distribute it over a proportionally larger surface, reducing their relative efficiency. Conversely, very small particles exhibit high efficiency but weak absolute absorption and poorly positioned resonances. Taken together, Figure S3 illustrates a coherent physical picture: intermediate-sized CDS nanoparticles with thin gold layers and moderate polymer coatings are optimal for achieving resonances within the first biological window while simultaneously maximizing expected directional photothermal efficiency. These findings provide a rational basis for narrowing the design space in subsequent stages, allowing for targeted optimization of structures intended for biomedical applications such as selective heating or photothermal therapy. However, a method or procedure is needed to rationally select the best candidate.

With these data, the goal is to maximize both absorption and efficiency to generate as much temperature as possible. While maximizing absorption is compatible with increasing the thickness of the polymer coating, simply by moving toward larger structures, another key objective of this work is to achieve efficient thermal directionality. Previous studies[4-6] have shown that the ability to direct heat depends on the presence of an insulating layer within the structure. Therefore, designs with significantly thick polymer layers will be explored. The simultaneous maximization of these three parameters imposes a trade-off between maximum absorbed energy and directional control, as increasing the polymer thickness reduces the particle's absorption efficiency. To determine the optimal structure based on a balanced maximization of these three parameters, the following trade-off function is defined:

$$\mathcal{F} = \bar{\sigma}_{abs}^{\gamma_1} \bar{Q}_{abs}^{\gamma_2} \bar{\delta}_p^{\gamma_3} \tag{S1}$$

where $\gamma_1, \gamma_2, \gamma_3 \in [1, 1.5]$ and

$$\bar{\sigma}_{abs} = \frac{\sigma_{abs}}{\max\{\sigma_{abs}\}} \tag{S2}$$

$$\bar{Q}_{abs} = \frac{Q_{abs}}{\max\{Q_{abs}\}} \tag{S3}$$

$$\bar{\delta}_p = \left(\frac{\delta_p}{\mathrm{r_c} + \delta_{Au} + \delta_p}\right)^2 \tag{S4}$$

All variables have been normalized so that their contribution to the trade-off function is dimensionless and as balanced as possible, given the contrasting nature of their absolute magnitudes ($\sigma_{abs} \ll Q_{abs} < \delta_p$). In all cases, the maxima of the absorption cross section and the absorption efficiency have been computed as the global maxima across the entire dataset. In this way, the trade-off function $\mathcal{F}$ is calculated considering all the exponents are equal to 1. The maximization process consisted in obtaining a two-dimensional data set for each $r_c$ containing the values for $\mathcal{F}$ as a function of $\delta_p$ and $\delta_{Au}$. Then the maximum value is stored for each $r_c$. Finally, the global maximum is obtained from a last maximization.

This information is shown in Figure S4. It can be observed that the compromise function yields a maximum for intermediate-sized structures, as expected. This behaviour arises because, in general, achieving higher absorption requires larger particles with minimal gold shell thicknesses, while the effect of the polymer layer remains marginal. However, when maximizing efficiency, larger structures are penalized, leaving intermediate-sized configurations as the most suitable candidates for reasonably thick polymer coatings. Therefore, for this particular set of exponents, the most appropriate configuration is: $(r_c, \delta_{Au}, \delta_p) = (36, 5, 21)$ nm.

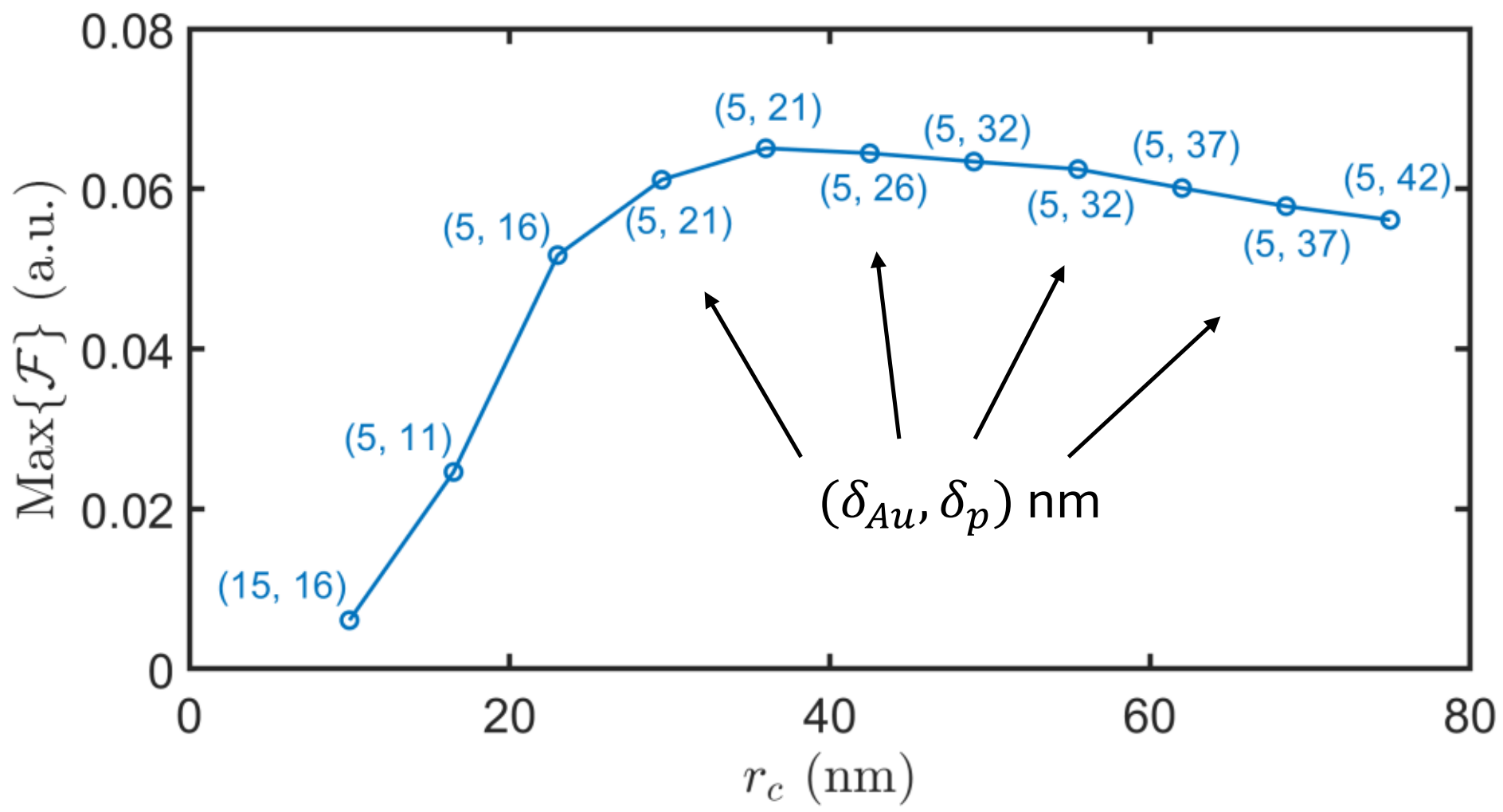

Figure S4. Local maxima of the trade-off function $\mathcal{F}$ for each core radius. Each ordered pair represents a vector with the corresponding gold ($\delta_{Au}$) and polymer ($\delta_p$) thicknesses respectively.

Besides, as shown in Figure S4, the trade-off function varies slowly for larger structures. This suggests that the influence of the different parameters may not be homogeneous. Although all quantities lie within the same order of magnitude, their dependence on the gold and polymer layer thicknesses can lead to uneven contributions, potentially masking the effect of one parameter over another. To mitigate this, three exponent values, $\gamma_1$, $\gamma_2$ and $\gamma_3$ are considered. The range of exponent values, from 1 to 1.5, has been deliberately chosen to prevent combinations that would cause one parameter to dominate the others, effectively reducing the compromise function to the maximization of a single quantity. By restricting the exponents to this moderate range, we ensure that all parameters contribute comparably, preserving the integrity of the trade-off function optimization.

The trade-off function is then computed over the entire dataset for every possible combination of exponents. In this way, for each triplet of exponents, the configuration that globally maximizes the trade-off function is identified. This approach not only yields a single global maximum but captures all possible global maxima associated with the different exponent sets. As a result, the selection of the optimal structure is reduced to a frequency-based analysis (Figure 3(a) in the main text), identifying a feasible range of structural parameters. This is consistent with the decision-making process required, since, given the passive nature of thermal transport processes, it is reasonable to expect that no single polymer thickness will be universally optimal.

To investigate the rotational stability of the ICDS under rotations, the peak absorption cross section is considered, as it is the parameter that describes the potential of a structure to produce heat. Therefore, the rotational stability is given in terms of the absorption sensitivity of each structure to rotational misalignment, i.e., misalignment between the particle iris and the incident electric field. We defined the absorption sensitivity metric as the maximum relative deviation in absorption across all sampled orientations for a given structural configuration $(\theta, \delta_p)$:

$$\mathcal{S}_\alpha = \left(1 - \frac{\min\{\sigma_{abs}(\alpha_i)\}_{i\in I}}{\max\{\sigma_{abs}(\alpha_i)\}_{i\in I}}\right) \cdot 100 \qquad (S5)$$

where $\sigma_{abs}(\alpha_i)$ is the absorption cross-section for orientation $\alpha_i$. This metric thus provides a quantitative estimation of the maximum change in absorption induced by particle reorientation for each structural configuration. Equation S5 shows that $\mathcal{S}_\alpha = 1$ implies full stability under rotations since it implies that $\min\{\sigma_{abs}(\alpha_i)\}_{i\in I} = \max\{\sigma_{abs}(\alpha_i)\}_{i\in I}$.

## II. Conductive thermal flux characterization.

The temperature-dependent thermal properties used in this work are shown in Figure S5. To clearly show the temperature dependence of those properties within the temperature range investigated here: 35 K temperature interval taking $T_0 = 293$ K as the initial temperature, each quantity was normalized to its value at 293 K. Besides, this enables a direct comparison between materials with very different absolute property values and shows their low relative variation in magnitude within the studied temperature interval. The corresponding values of thermal conductivity, specific heat and mass at the initial temperature are given in Table S1.

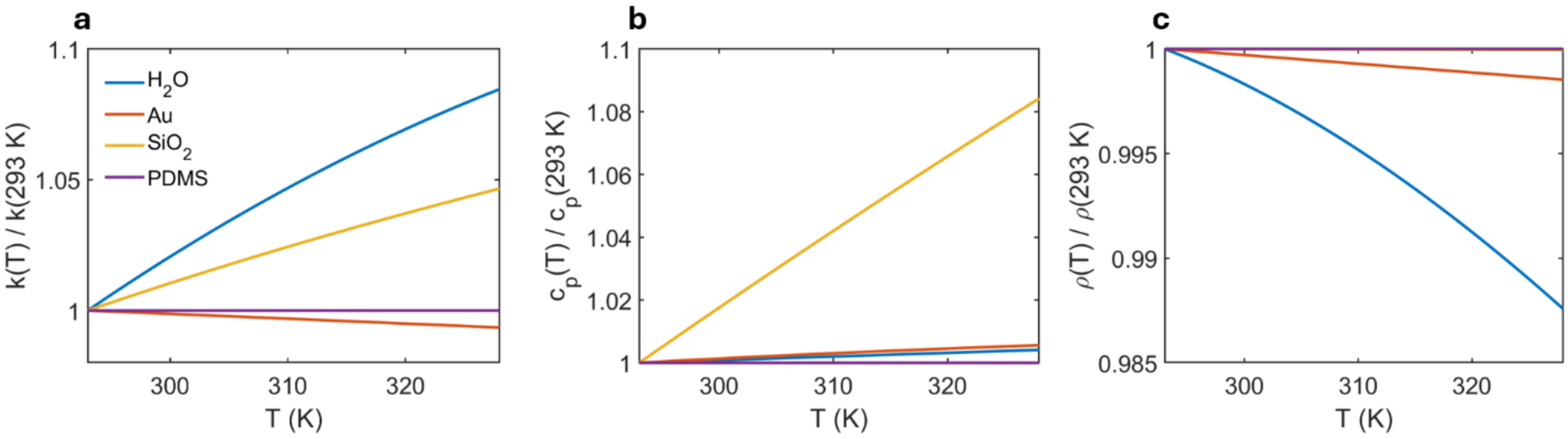

Figure S5. Relative variation of the thermal properties used in the simulations over the temperature range 293–328 K for water (blue), gold (orange), silica (yellow) and PDMS (purple). (a) thermal conductivity, (b) specific heat, and (c) mass density. Each property is normalized to its corresponding value at the initial temperature, $T_0 = 293$K.

| Material | $k$[W m$^{-1}$ K$^{-1}$] | $c_p$[J kg$^{-1}$ K$^{-1}$] | $\rho$[kg m$^{-3}$] |
|---|---|---|---|
| $H_2O$ | 0.60 | 4180 | 998 |
| Au | 317.7 | 128.6 | 19282 |
| $SiO_2$ | 1.37 | 704 | 2220 |
| PDMS | 0.16 | 1460 | 970 |

Table S1. Thermal properties of water, gold, silica and PDMS at $T = 293$ K.

One of the main objectives of this manuscript is not only to quantify the amount of thermal energy delivered to each half-space, but also to assess the degree of directionality or focusing of that energy as a function of the configuration. To characterize the degree of directionality in COMSOL, the conductive heat flux is integrated over a spherical surface, concentric with the nanoparticle under analysis. Using this same surface, the same quantity is iteratively integrated over a spherical cap defined by a solid angle Ω, which subtends a semi-aperture angle $\alpha$ such that:

$$\Omega(\alpha) = 2\pi R^2(1 - \cos(\alpha)) \tag{S6}$$

By dividing the heat flux crossing the spherical cap by the total flux over the full spherical surface, one obtains the fraction of thermal energy generated by the particle that is emitted into a given region of space. This quantity provides a direct measure of the degree of thermal directionality. This procedure yields a value for each chosen semi-aperture angle $\alpha$. Accordingly, we define the *directionality angles* as the set $\{\alpha_i\}_{i\in I}$ such that the fraction of energy crossing the corresponding spherical caps $\{\Omega_i\}_{i\in I}$, are $\{\eta_i\}_{i\in I}$ respectively.

To validate and illustrate the suitability of this approach, we perform the analysis in the paradigmatic case of a gold nanosphere embedded in a homogeneous aqueous environment under continuous illumination. In this scenario, the heat flux is expected to be purely radial and isotropic[7], so that in spherical coordinates the vector field can be written as: $\vec{\Phi}(r,\theta,\phi) = (\Phi(r),0,0)$. Given that the outward normal to the spherical integration surface is $\vec{n} = \vec{e}_r$, in this coordinate system, it follows that:

$$\eta(\alpha) = \frac{\iint_{\Omega(\alpha)} \vec{\Phi} \cdot \vec{n}\, dS}{\iint_{\mathbb{S}^2} \vec{\Phi} \cdot \vec{n}\, dS} = \frac{1}{2}(1 - \cos(\alpha)) \qquad (S7)$$

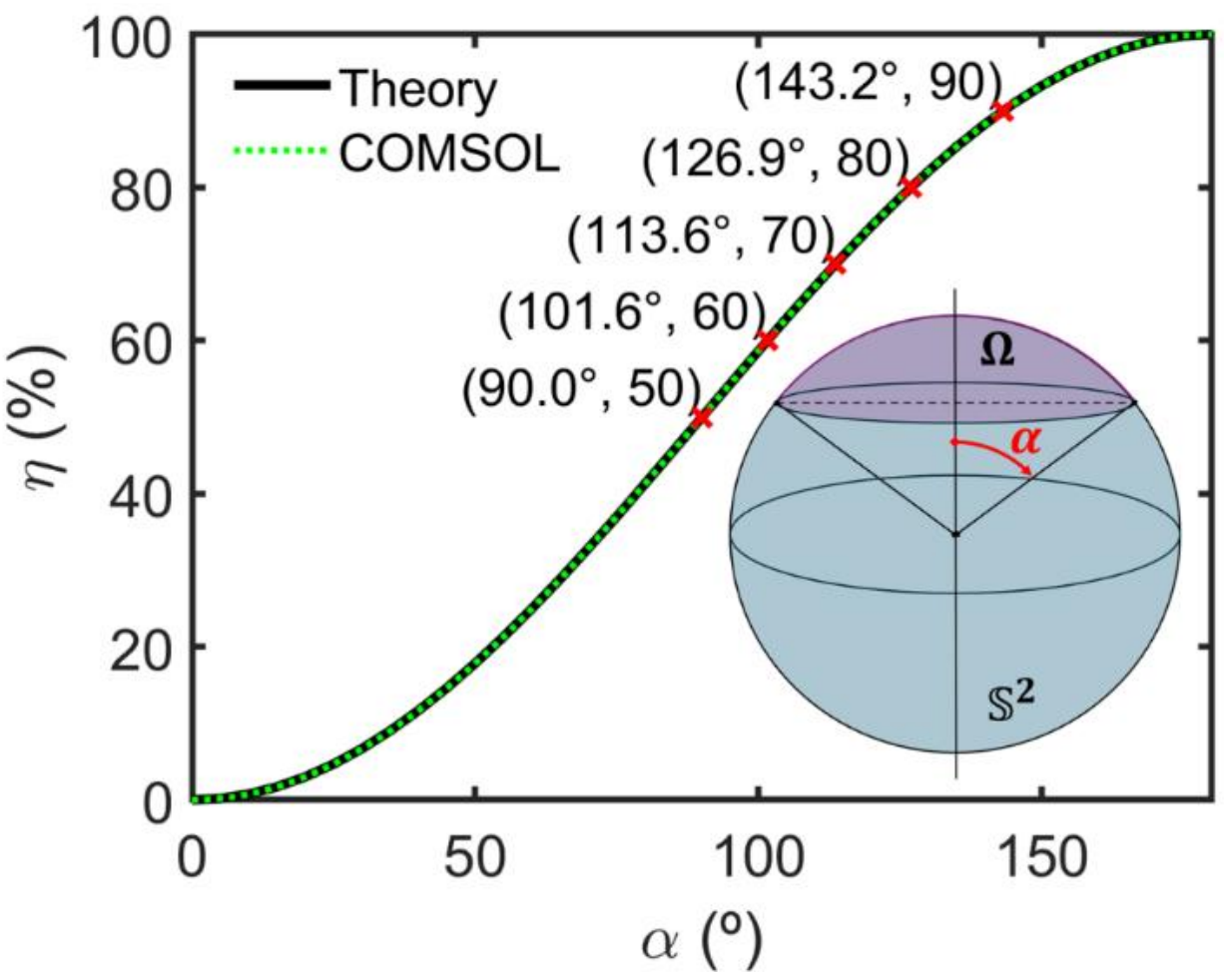

Figure S6. Power fraction $\eta(\alpha)$ crossing a spherical cap of semi-aperture angle $\alpha$, computed for a homogeneous gold sphere in water under continuous illumination. The analytical prediction (black curve) is in excellent agreement with COMSOL simulations (green dashed line). Red crosses indicate specific threshold values, with each pair $(\alpha, \eta)$ marked on the curve. Notably, an aperture angle of 90° encloses exactly 50% of the emitted power, while capturing 90% of the thermal energy requires $\alpha \approx 143°$. The inset illustrates the integration geometry, where $\Omega$ denotes the cap subtended by $\alpha$ over the spherical surface $\mathbb{S}^2$.

Figure S6 shows the fraction of thermal power crossing a spherical cap as a function of the semi-aperture angle defining the cap, in the case of a homogeneous gold sphere. It can be observed that for an angle of 90°, which corresponds to the entire upper half-space, the particle delivers exactly 50% of the generated energy. To enclose 90% of the total thermal power, the aperture must be extended beyond 140°. This result highlights the absence of directionality in this reference structure. When combined with the temperature map, this analysis provides an excellent baseline for comparing the directional behaviour and degree of thermal focusing in more complex architectures. Since we are interested in the directionality towards the upper hemisphere, the value $\eta = 50\%$ will be typically taken as a reference since in the symmetric case corresponds to the half of the total generated power.

Even though this metric is useful to analyse the asymmetry of the thermal flux generated by a Janus particle, it introduces a dependency on distance, i.e., the radius of the spherical integration surface. To test the sensitivity of our results to this parameter, we have performed a variational study for the reference case where $\theta = 70°$, evaluating the hemispherical power fraction at control-surface distances $\Delta = 5,10,15,20,\text{and } 25\text{nm}$.

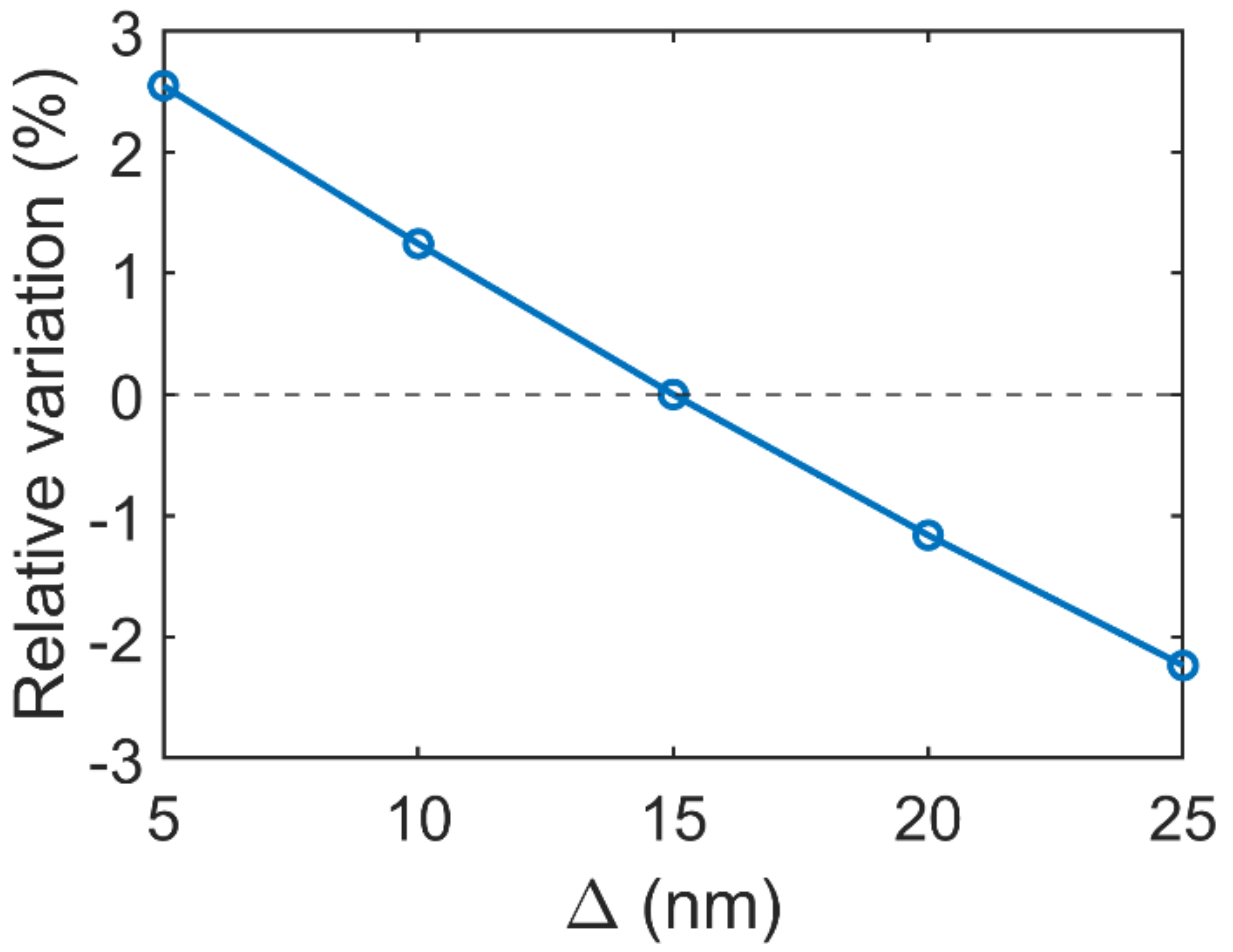

Figure S7. Relative variation of the hemispherical thermal power fraction as a function of the distance $\Delta$ between the exposed nanoparticle surface and the spherical integration surface, evaluated for the reference ICDS configuration with $\theta = 70°$. The value at $\Delta = 15$ nm is taken as the reference.

The value obtained at $\Delta = 15$nm was taken as the reference, and the relative variation of the integrated power fraction was calculated for the remaining distances. This interval was selected to consider separation lengths compatible with typical nanoparticle biofunctionalization layers and nanoparticle–biological-interface distances. As shown in Figure S7, the relative variation remains below approximately $2.6\%$ over the range considered. Therefore, although the metric is formally distance dependent, its sensitivity is weak within biologically relevant distances, supporting the assumption that the main conclusions regarding thermal directionality remain unchanged over this range.

**III. Energy density intensification and gain parameter definition.**

a) Physical motivation.

The asymmetric structures under consideration redirect part of the thermal flux toward the upper hemisphere, thereby concentrating the emitted energy over a smaller spherical surface compared to a symmetric configuration. To quantify the resulting local intensification of surface energy density, we define a relative gain parameter $G$ that, for each $\eta$, compares the spherical area covered by that fraction of energy in the studied structure to that of the symmetric case.

b) Definition of $G$.

As seen in Figure S6, the generated energy fraction is defined as the amount of energy per unit time (i.e., power) that passes through a spherical surface subtended by a solid angle $\Omega(\beta)$, normalized by the total power emitted by the structure. This quantity captures how much of the emitted energy is directionally distributed within a given angular region and serves as a basis for evaluating thermal redirection and focusing effects.

$$\eta(\beta) = \frac{\iint_{\Omega(\beta)} \vec{\Phi} \cdot \vec{n}\, dS}{\iint_{\mathbb{S}^2} \vec{\Phi} \cdot \vec{n}\, dS} = \frac{P(\beta)}{P_T}$$

In this way, for each energy fraction, one can associate a spherical surface area defined by the solid angle $\Omega(\beta)$, which in turn allows that energy level to be identified with a corresponding semi-angle $\beta$ in a well-defined mathematical manner. The area of this spherical surface is expressed as:

$$A(\beta) = 2\pi R^2(1 - \cos(\beta)) \tag{S8}$$

where $R$ is the radius of the control sphere surface through which the energy flux is evaluated. This expression for the area can be further particularized within our context. Specifically, the area of the spherical surface subtended by the energy fraction $\eta$ for a structure with an aperture angle $\theta$ is given by:

$$A(\eta, \theta) = 2\pi R^2\big(1 - \cos\big(\beta_\theta(\eta)\big)\big) \tag{S9}$$

Thus, for a fixed energy fraction $\eta$, the power density crossing the spherical control surface can be computed as:

$$I(\eta, \theta) = \frac{\eta \cdot P_{tot}}{A(\eta, \theta)} = \frac{\eta \cdot P_{tot}}{2\pi R^2\big(1 - \cos\big(\beta_\theta(\eta)\big)\big)} \tag{S10}$$

where $P_{tot}$ is the total power generated by the structure. This expression quantifies how concentrated the emitted energy is on the spherical surface, serving as the basis for defining the relative gain. Given a fixed value of $\eta$, there exists a corresponding angle $\beta$ that defines the spherical cap enclosing that energy level. To assess how the energy density on the control surface becomes enhanced, we compute the angle $\beta_0(\eta)$ that defines the spherical cap required to enclose the same energy fraction $\eta$ in the **symmetric case** ($\theta = 0°$). The associated area is denoted as $A(\eta, 0)$. Accordingly, the power density in the symmetric case can be expressed as:

$$I(\eta, 0) = \frac{\eta \cdot P_{tot}}{A(\eta, 0)} = \frac{\eta \cdot P_{tot}}{2\pi R^2\big(1 - \cos\big(\beta_0(\eta)\big)\big)} \tag{S11}$$

It is worth noting that, since the objective is to assess the **energy density**, the absolute magnitude of the total emitted power is not of primary importance. For this reason, we assume a symmetric reference case that emits the same total power as the asymmetric structure, within the same spatial domain. Under this assumption, the **gain** is defined by comparing the power density of the asymmetric configuration with that of the symmetric case, leading to:

$$G(\eta, \theta) = \left(\frac{I(\eta, \theta)}{I(\eta, 0)} - 1\right) \cdot 100 = \left(\frac{1 - \cos\big(\beta_0(\eta)\big)}{1 - \cos\big(\beta_\theta(\eta)\big)} - 1\right) \cdot 100\ [\%] \tag{S12}$$

which expresses the relative enhancement of energy density induced by the asymmetry, at a given energy level $\eta$ for the aperture angle $\theta$, in percentage terms. Since any degree of structural asymmetry induces thermal asymmetry, the ratio $I(\eta, \theta)/I(\eta, 0)$ is always greater than 1. Therefore, for the gain $G(\eta, \theta)$ to represent a **percentage increase** in energy density, the unit value is subtracted from the ratio. This allows for a direct and visual interpretation of the gain as:

- $G > 0$: the same energy fraction $\eta$ is confined within a smaller spherical cap, thereby increasing the energy density crossing the surface.
- $G = 0$: no intensification occurs, which corresponds to a fully symmetric case.
- The larger the value of $G$, the stronger the thermal focusing effect.

c) Physical interpretation of the gain metric.

A particularly illustrative case is that of $\eta = 0.5$, which corresponds to the fraction of energy directed into the upper hemisphere ($\beta = 90°$) in a fully symmetric configuration ($\theta = 0°$). In such a symmetric case, the energy is uniformly emitted in all directions, so precisely half of the total energy flows through the upper half-space. This sets a reference angular threshold: $\beta_0(0.5) = 90°$, and the corresponding area is that of a hemisphere.

When structural asymmetry is introduced, the same energy fraction $\eta = 0.5$ becomes concentrated into a smaller solid angle, i.e., a cap of radius $\beta_\theta(0.5) < 90°$. This angular compression implies that the same amount of energy crosses a reduced surface, increasing the local energy density. The relative gain $G(\eta, \theta)$ thus quantifies this enhancement.
For instance, if a given structure yields $G = 30\%$ at $\eta = 0.5$, this means that the half of the total emitted energy is now confined within a surface whose area is 30% smaller than that of the hemisphere. From the gain expression:

$$G(\eta, \theta) = \left( \frac{1 - \cos\big(\beta_\theta(\eta)\big)}{1 - \cos\big(\beta_0(\eta)\big)} - 1 \right) \cdot 100 \qquad (S13)$$

and knowing that $\cos(\beta_0) = cos(90°) = 0$, we find that $1 - \cos(\beta_\theta) = 0.7$, so $\cos(\beta_\theta) = 0.3$, which gives $\beta_\theta \approx 72.5°$. In this case, 50% of the energy is emitted within a cone of only 72.5°, compared to 90° in the symmetric case.
For a stronger effect, say $G = 50$ %, we have $1 - \cos(\beta_\theta) = 0.5$, so $\cos(\beta_\theta) = 0.5$, yielding $\beta_\theta = 60°$. This indicates that the same amount of energy is now emitted within just one quarter of the spherical surface, a strong indication of directional thermal emission.
This metric thus offers a direct and quantitative way to evaluate how effectively a structure redirects and concentrates thermal energy, providing insight into both the angular confinement and surface densification induced by asymmetry.

**IV.-Calculation of the relaxation time.**

The pulse duration under investigation is defined based on the computed transient thermal response. The system's peak temperature is subsequently fitted using the following model[7]:

$$m(t) = A\left(1 - \mathrm{erf}\left(\frac{B}{\sqrt{t}}\right)\right) \qquad (S14)$$

where $A = T_\infty$ is a fitting constant associated with the steady-state temperature, while $B^2 = R/4D_s$ represents the squared fitting constant corresponding to the characteristic heating time. According to this expression, the parameter $B^2$ can be interpreted as the time at which the temperature reaches approximately 16% of the stationary value. However, in this work, the relevant timescale is taken as the time required to reach 84% of the stationary temperature. This matches the asymptotic growth properties of the error function.

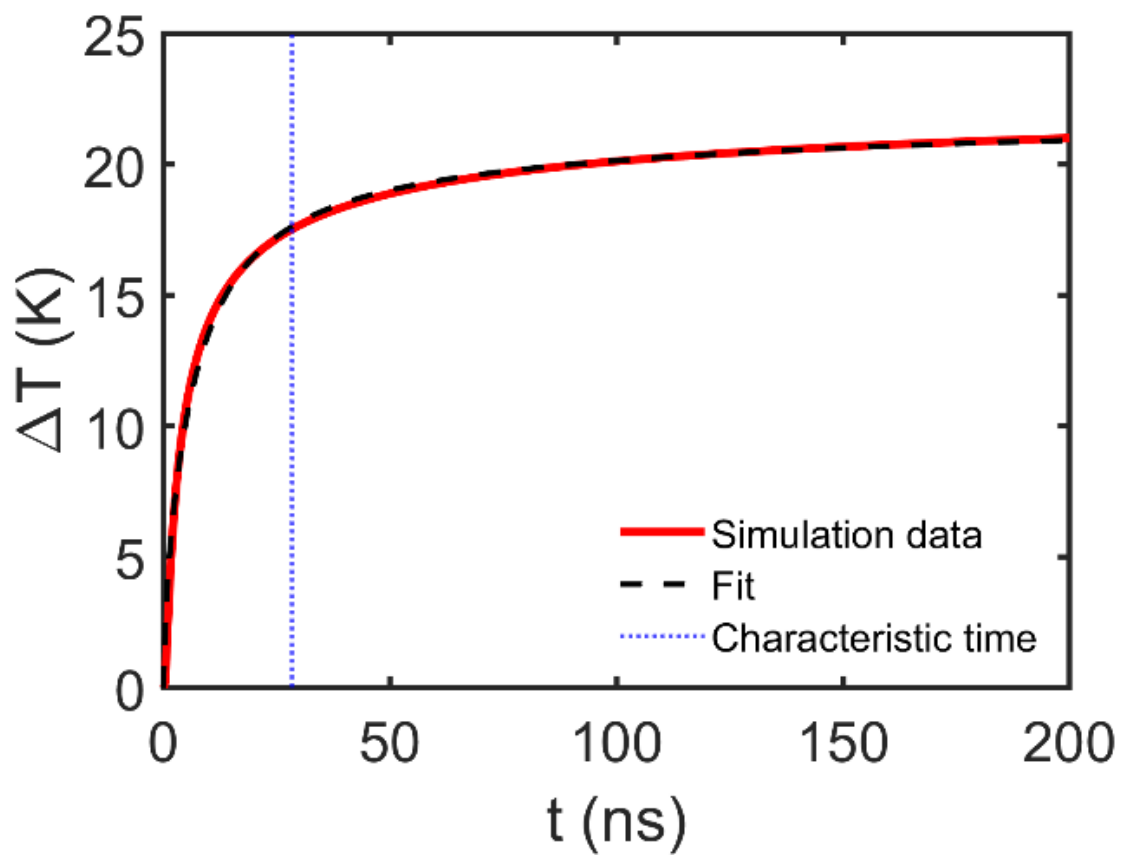

Figure S8. Temperature rise of a CDS defined by $(r_c,\ \delta_{Au}, \delta_p, \theta) = (36\ \mathrm{nm}, 5\ \mathrm{nm}, 40\ \mathrm{nm}, 70^0)$. The red and the black dashed lines correspond to the simulated and fitted temperatures. The blue vertical line stands for the time for which the structure reaches 84% of the stationary temperature.

The fitting is shown in Figure S8. It can be seen that the model fits very well the simulated data. The relative fitting error $\sigma$ is calculated as the mean of the sum of squared residuals over the stationary temperature: $\sigma \approx 0.2\%$. Based on this fitting, a characteristic heating time of approximately 30 ns is obtained under the present conditions.